\documentclass[12pt,twoside]{article}

\usepackage[usenames]{color}
\usepackage{lscape}
\usepackage{amssymb}
\usepackage{rotating}
\usepackage[T1]{fontenc}
\usepackage{ae}
\usepackage{aecompl}
\usepackage{pifont}
\usepackage{graphicx}
\usepackage{fancyhdr}
\usepackage{amsmath}
\usepackage{times}
\usepackage{floatflt}
\usepackage{hyperref}

\def\pa{\partial}
\def\nn{\nonumber \\}
\def\ov{\overline}
\def\br{&\!\!\!\!}

\newlength{\dinwidth}
\newlength{\dinmargin}
\begin{document}

\thispagestyle{empty}
\begin{flushright}
IFT--08--11\\
\end{flushright}

\vspace*{15mm}

\begin{center}
{\Large\bf
Volume modulus inflection point inflation
\\[6pt]
and the gravitino mass problem
}
\vspace*{5mm}
\end{center}
\vspace*{5mm} \noindent
\vskip 0.5cm
\centerline{\bf
Marcin Badziak\footnote[1]{mbadziak@fuw.edu.pl}
and Marek Olechowski\footnote[2]{Marek.Olechowski@fuw.edu.pl}
}
\vskip 5mm
\centerline{\em Institute of Theoretical Physics,
University of Warsaw}
\centerline{\em ul.\ Ho\.za 69, PL--00--681 Warsaw, Poland}

\vskip 15mm

\centerline{\bf Abstract}
\vskip 3mm
Several models of inflection point inflation with the volume
modulus as the inflaton are investigated. Non-perturbative
superpotentials containing two gaugino condensation terms or
one such term with threshold corrections are considered.
It is shown that the gravitino mass may be much smaller than
the Hubble scale during inflation if at least one of the
non-perturbative terms has a positive exponent.
Higher order corrections to the K\"ahler potential have to be
taken into account in such models.
Those corrections are used to stabilize the potential in
the axion direction in the vicinity of the inflection point.
Models with only negative exponents require uplifting and
in consequence have the supersymmetry breaking scale higher
than the inflation scale. Fine-tuning of parameters and initial
conditions is analyzed in some detail for both types of models.
It is found that fine-tuning of parameters
in models with heavy gravitino is much stronger than in models
with light gravitino.
It is shown that recently proposed time dependent potentials
can provide a solution to the problem of the initial conditions
only in models with heavy gravitino. Such potentials can not be
used to relax fine tuning of parameters in any model because
this would lead to values of the spectral index well outside
the experimental bounds.

\newpage

\section{Introduction}

Mechanism of inflation has many phenomenologically desired properties.
In the last couple of years a lot of attention was focused on
inflationary model building motivated by the string theory.
This is mainly due to a significant progress in moduli stabilization.
Fluxes were used in \cite{Giddings} to stabilize the dilaton and
the complex structure moduli. In order to stabilize the remaining
volume modulus, such fluxes were supplemented by non-perturbative
contributions to the superpotential and uplifting of the potential
in the KKLT scenario \cite{kklt}.
Since the moduli fields are necessary ingredients of 4D supergravity
(obtained as the low energy limit of the compactified string theory)
it is very natural to ask whether they can play the role of the inflaton.
With the advent of racetrack inflation \cite{racetrack} and its
generalizations \cite{westphal}-\cite{betrace} it turned out that
the volume modulus may be a good candidate for the
inflaton. Other models of inflation driven by moduli fields were
investigated e.g.\ in \cite{lalak,conlon}.
In this paper we concentrate on the class of models
in which effectively only one modulus - the volume K\"ahler modulus
- plays an important role in inflation.

One of the generic properties of inflationary models based on the
KKLT mechanism is the incompatibility between a high scale of inflation
and a low scale of supersymmetry (SUSY) breaking. It was pointed out in
\cite{KaLi} that the gravitino mass is typically bigger than the Hubble
scale during inflation.
The authors of \cite{KaLi} noticed that this problem could be avoided
if inflation ends in a SUSY (near) Minkowski
minimum\footnote
{In \cite{Pillado} it was shown that any SUSY Minkowski stationary point
is stable up to possible flat directions.
}.
However, even if such minimum exists it is still very hard to construct
a model of inflation ending in that minimum. One of the main problems is
that the slow-roll parameter $\eta$ is necessarily smaller than $-2/3$ in
models with the tree level K\"ahler potential and arbitrary
superpotentials \cite{bo}. The situation can be improved if subleading
corrections to the K\"ahler potential are taken into account.
A model using those corrections was proposed in \cite{bo}.
However, that model is quite complicated: it involves three
non-perturbative gaugino condensation terms in the superpotential
and requires more fine-tuning of parameters as compared to typical
inflationary models.

It was found recently \cite{lvgrav} that the Hubble scale during
inflation is related to the gravitino mass also in the large volume
compactification scenario . In this class of models typically
$m_{3/2}^{3/2}\geq HM_{\rm Pl}^{1/2}$ which makes low-energy SUSY breaking
even more problematic. A possible solution to this problem was
proposed by constructing a model in which a Minkowski minimum
occurs at exponentially large volume with exponentially suppressed
gravitino mass \cite{lvgrav}.
However, this model suffers from the overshooting
problem and requires significant amount of fine-tuning. Therefore,
the problem with large gravitino mass is generic in string
inflationary models and not easy to overcome.

The relation between the inflationary Hubble scale and the gravitino
mass does not have to be considered problematic if we do not insist
on a high scale of inflation. However, it is quite difficult to 
construct models with a low scale of inflation. Such models have been
proposed \cite{German,Allahverdi} but usually they require extremely
small value of the slow-roll parameter $\epsilon$ which is obtained
by severe fine-tuning of parameters \cite{turzynski}. In the present
paper we concentrate on more typical models with a high scale of
inflation.

The gravitino should not be very heavy if supersymmetry is to solve 
the hierarchy problem. On the other hand, it is well known that 
small gravitino mass is typically incompatible with the primordial 
nucleosynthesis or leads to the overclosure of the Universe 
(if it is stable). This is the famous gravitino problem 
\cite{Weinberg,Khlopov}. Investigation of that problem is 
beyond the scope of this paper. We assume that it can be solved 
by another sector of the theory.

In this paper we investigate models in which inflation takes place
when the inflaton is close to an inflection point of the potential.
Recently, models of this type gained a lot of attention in the
context of brane inflation \cite{KrPa}-\cite{Panda} as well as in
the context of MSSM inflation \cite{Allahverdi,Sanchez}.
Here, we focus on inflection point inflation driven by the volume
modulus. This is motivated also by the fact that a mechanism to solve
the problem of the initial conditions for the inflaton in this type
of models was proposed \cite{Itzhaki}.
We present a few examples in which inflation starts near
an inflection point and ends in a SUSY (near) Minkowski minimum.
The fine-tuning of the parameters in those models is substantially
weaker than in our previous model with three gaugino condensation
terms \cite{bo}. We analyze also models with uplifting which do not
accommodate a TeV-scale gravitino mass and compare their features to
the models with a SUSY (near) Minkowski minimum.

Very important ingredients of our models are positive exponents in
non-perturbative terms in the superpotential. Such terms may
occur if the gauge kinetic function is a linear combination of
closed string moduli
\cite{Abe2}-\cite{Abe4}:
\begin{equation}
\label{gaugekin}
f=w_SS+w_TT \,.
\end{equation}
If gaugino condensation occurs, this results in the following
non-perturbative contribution to the superpotential:
\begin{equation}
W_{\rm np}=Be^{-\frac{2\pi}{N}\left(w_SS+w_TT\right)} \,,
\end{equation}
where $N$ is the rank of the hidden sector gauge group SU($N$).
In the leading order approximation, the prefactor $B$ is just a
constant. In general, it becomes a function of the moduli when
the threshold corrections are taken into account
\cite{gaugino_corr1,gaugino_corr2}. We will consider
both situations: models with constant prefactors and models with
moduli-dependent ones.
Assuming that the dilaton $S$ is stabilized by
fluxes\footnote
{See \cite{Choi}-\cite{Ach} for analysis on validity of this assumption.
}
with the mass much bigger than the mass of $T$
we obtain the effective superpotential for the volume modulus
\begin{equation}
\label{Weff}
W^{\rm eff}_{\rm np}=B^{\rm eff}e^{bT} \,,
\end{equation}
where $B^{\rm eff}=Be^{-\frac{2\pi}{N}w_S\left\langle S\right\rangle}$
and $b=-\frac{2\pi}{N}w_T$.
In some models it may happen that $w_T$ is negative,
and the exponent $b$ is positive.
We will use such non-perturbative terms with positive exponents
in models \ref{posneg}, \ref{twopos} and \ref{posthr}
(to name the models presented in this paper we use
numbers of subsections in which they are described
i.e.\ model \ref{posneg} is investigated in subsection \ref{posneg} etc.).
Of course, the real part of
the gauge kinetic function has to be positive since it determines
the physical coupling constant. Therefore, a theory with the effective
superpotential (\ref{Weff}) is reliable only in the region
where $w_S\left\langle{\rm Re} S\right\rangle+w_T{\rm Re} T>0$.
Throughout this paper we will assume that this condition is satisfied.
Observe that, for a large vacuum expectation value of the dilaton,
it is quite natural that the effective prefactor $B^{\rm eff}$
can be much smaller than the initial prefactor $B$.

We should stress that gauge kinetic functions that
involve more than one modulus are quite common in string theory, as was
pointed out in \cite{Abe2}-\cite{Abe4}. Furthermore, in some string
theories $w_T$ can be negative leading to positive exponents in the
superpotential. This is the case for example in heterotic M-theory
\cite{Lukas}. In the context of type IIB string theory such gauge
kinetic functions can occur on magnetized D9-branes
\cite{Cascales,Marchesano}.

Models with positive exponents were explored in a series of papers
\cite{Abe2}-\cite{Abe4}, where it was noticed that such forms of
the superpotential may have many interesting cosmological applications.
An example of such application was presented in \cite{Abe4} where
a racetrack-type inflation model \cite{racetrack} was considered.
It was demonstrated that adding to the superpotential a term with a positive
exponent may solve the overshooting problem by increasing the height
of the barrier separating the vacuum from the run-away region.
Superpotentials with positive exponents were investigated also
in the context of moduli stabilization \cite{Dudas,Curio}.

The rest of this paper is organized as follows.
In section \ref{2condensates} we investigate models with two gaugino
condensates in the superpotential for all possible combinations
of the signs of the exponents. In some cases inflation ending
in a SUSY Minkowski vacuum is possible. In other cases we have to add
appropriate uplifting terms to the potential obtaining inflationary
models with a heavy gravitino.
In section \ref{1condensatethr} we study single gaugino condensation
models with a modulus-dependent prefactor of the exponential term in
the superpotential (resulting from threshold corrections to gauge
kinetic functions). For simplicity we focus on a case in which
such prefactor depends
linearly
on the volume modulus.
We investigate separately the case of positive and negative exponent.
Phenomenological implications of all our models are shortly
discussed in section \ref{exp}.
In section \ref{timedepend} we investigate time dependent potentials
proposed in \cite{Itzhaki} and discuss their impact on the fine-tuning
of our models. We show that such potentials can provide solution to
the problem of initial conditions in models with uplifting.
We consider a simple toy-model to show that such potentials can not
reduce fine-tuning of the parameters because they lead to the
spectral index incompatible with the present data. In section
\ref{discussion} we summarize and compare the main features of our models.


\section{Models with double gaugino condensation}
\label{2condensates}

The F-term potential in supergravity is given, in terms of the
superpotential $W$ and the K\"ahler potential $K$, by the following formula:
\begin{equation}
        \label{potentialWK}
        V=e^K\left(K^{i\ov{j}}D_i W\ov{D_j W}-3\left|W\right|^2\right) \ ,
\end{equation}
where $D_i W=\pa_i W+W\pa_i K$.

In this section we consider models with two gaugino condensation
terms in the superpotential:
\begin{equation}
\label{sup}
W=A+Ce^{cT}+De^{dT}
\,.
\end{equation}
We assume, for simplicity, that all parameters are real.
The parameters $C$ and $D$ may be effective in a sense
that they may depend on the vev of the dilaton as explained in the
introduction (we drop the superscript ``eff'' for further convenience).
We are interested in SUSY Minkowski minima which occur if
two conditions: $W=0$, $\partial_TW=0$, are fulfilled simultaneously.
This may happen only when the parameters in the superpotential
(\ref{sup}) fulfill the following condition:
\begin{equation}
A=
-C\left(-\frac{cC}{dD}\right)^{\frac{c}{d-c}}
-D\left(-\frac{cC}{dD}\right)^{\frac{d}{d-c}}
.
\label{A}
\end{equation}
Then, the SUSY Minkowski minimum is located at
\begin{equation}
T_{\rm Mink}=t_{\rm Mink}+i\tau_{\rm Mink}
=\frac{1}{d-c}\ln\left({-\frac{cC}{dD}}\right)
.
\label{T_Mink}
\end{equation}
The gravitino in such a minimum (unbroken SUSY) is exactly massless.
In order to have a non-zero (but small) gravitino mass we should
relax (slightly) the fine-tuning condition (\ref{A}) for the
superpotential parameters. Then, the minimum of the potential moves
a little bit away from the position given by eq.\ (\ref{T_Mink}).
It is still supersymmetric but becomes anti de Sitter (AdS).
Some mechanism is necessary to uplift it again to (almost) vanishing
energy (this is the usual cosmological constant problem which
we are not going to discuss). However, this uplifting is much smaller
than in typical KKLT-type models. So, we will neglect it and consider
models in which the superpotential parameters are (approximately)
fine-tuned as in eq.\ (\ref{A}) and
with inflation ending in SUSY (near) Minkowski minima located
(approximately) at $T$ given by (\ref{T_Mink}).

In the present work we concentrate on models in which the real
part of $T$ plays the role of the inflaton. Such inflation can
be realized when the (near) Minkowski minimum and the inflection
(or saddle) point, at which inflation starts, are both located at
$\tau=0$. The real part of the volume modulus at the Minkowski minimum,
$t_{\rm Mink}$, must be positive.
It follows from eq.\ (\ref{T_Mink}) that $T_{\rm Mink}$ is real
and positive if
\begin{equation}
cCdD<0
\,,
\qquad\qquad
(d-c)|cC|>(d-c)|dD|
\,.
\label{cCdD}
\end{equation}
Note that the first of the above conditions is also sufficient to
fulfill the condition (\ref{A})
with all parameters real, as we have assumed.

The leading order K\"ahler potential has the following form:
\begin{equation}
\label{leadK}
K=-3\ln(T+\ov{T}) \ .
\end{equation}
It is easy to check that for an arbitrary superpotential with real
parameters the F-term potential has some vanishing derivatives for
$\tau=0$: $\left(\pa V/\pa\tau\right)|_{\tau=0}=0$,
$\left(\pa^2 V/\pa t\pa\tau\right)|_{\tau=0}=0$.
Therefore, $\eta$ matrix is diagonal at all stationary points
satisfying $\tau=0$.
However, the sign of $\left(\pa^2 V/\pa\tau^2\right)|_{\tau=0}$
may change with $t$ in a way depending on details of
a given model. Successful inflation in $t$ direction requires
$\left(\pa^2 V/\pa\tau^2\right)|_{\tau=0}$, which in our case is
proportional to $\eta_\tau^\tau$, to be positive for any $t$
between the inflationary inflection (or saddle) point
and the Minkowski minimum, where inflation is supposed to end.
In addition, one of the slow roll conditions requires that
during inflation $|\eta_t^t|\ll1$.
It was shown in \cite{bo} (see also \cite{deAl,covi2}) that for
the tree-level K\"ahler potential, the trace of the $\eta$ matrix
equals $-4/3$ at any non-supersymmetric stationary point.
At any such point $\eta_t^t=-4/3-\eta_\tau^\tau<-4/3$,
and the slow roll condition $|\eta_t^t|\ll1$ is strongly
violated\footnote
{
The same problem from another point of view:
at (close to) any non-supersymmetric stationary point
with $|\eta_t^t|\ll1$ we find $\eta_\tau^\tau\approx-4/3$,
and the potential is strongly unstable in the $\tau$ direction.
}.
Successful inflation can not start near any
non-supersymmetric stationary point in models with the minimal
K\"ahler potential (\ref{leadK}). The same arguments apply
also to models in which one tries to start inflation near
an inflection point. The reason is that a flat inflection
point is typically either close to a flat saddle point or
may be transformed to a flat saddle point by a very small
change in the model parameters.

The above problem of too big negative $\eta_t^t$
can be solved by including some corrections
to the K\"ahler potential. That is why
we use the K\"ahler potential with the leading $\alpha'$ and
string loop corrections \cite{cn}-\cite{gersdorff}:
\begin{equation}
K=-3\ln(T+\ov{T})-\frac{\xi_{\alpha'}}{(T+\ov{T})^{3/2}}
-\frac{\xi_{\rm loop}}{(T+\ov{T})^2} \ .
\label{Kcorr}
\end{equation}
It is important to stress that the inclusion of the corrections to
the K\"ahler potential implies that the potential, at this approximation,
is singular at some value of $t=t_\infty$ at which the inverse of the
K\"ahler metric which enters the potential,
\begin{equation}
K^{T\ov{T}}=\frac{4t^2/3}{1-\frac{5}{4}\frac{\xi_{\alpha'}}{(2t)^{3/2}}
-2\frac{\xi_{\rm loop}}{(2t)^2}}
\,,
\end{equation}
is singular. As was shown in \cite{bo}, the coefficients
$\xi_{\alpha'}$ and $\xi_{\rm loop}$ should be positive
to make the trace of the $\eta$ matrix positive. On the other hand,
the K\"ahler metric has to be positive definite, so the corrections
should not be too large. Therefore, the region of the potential crucial
for inflation should be far away from the singularity $t_{\infty}$
in order to have the corrections under control.

The potential for the modulus $t$ is of the following form:
\begin{eqnarray}
\label{potdouble}
V=\frac{1}{6t^2}
&&\!\!\!\!\!\!\!\!
\left[CD \left( 2cdt-3c-3d \right)
{e^{ \left( c+d \right)t }}+{C}^{2}c \left( ct-3 \right) {e^{2ct}}
\right.
\nn[4pt]
&&\,\,\,\,\,\,\,\,
+ \left.
{D}^{2}d \left( dt-3 \right) {e^{2dt}}-3A \left( Cc{e^{ct}}+
 D  d{e^{dt}} \right) \right]+ \ldots
\,,
\end{eqnarray}
where we set $\tau=0$ and the ellipsis stands for the terms involving
$\xi_{\alpha'}$ and $\xi_{\rm loop}$ coming from the corrections to
the K\"ahler potential. Those corrections are necessary to ensure the
stability of the trajectory $\tau=0$ in the $\tau$ direction.
We do not present explicitly the potential with the corrections since
it is very complicated. Moreover, the corrections does not change
qualitatively the potential in the $t$ direction except the fact
that the singularity appears, as discussed before.

The potential in models with two gaugino condensates has an
interesting feature: for some regions of the parameter space,
there are two minima in the $t$ direction for $\tau=0$.
One of them may be a SUSY (near) Minkowski minimum.
When both exponents $c$ and $d$ are negative, like in the
Kallosh-Linde (KL) model \cite{KaLi},
the second minimum is also supersymmetric and has negative energy.
In models with (at least) one positive exponent,
the second minimum is non-supersymmetric and may have positive energy.
We will see that it is crucial for inflation whether this additional
minimum is supersymmetric or not.

In the following we consider in turn models with all
possible sign assignments for the exponents $c$ and $d$.


\subsection{One positive and one negative exponent}
\label{posneg}

\begin{figure}[t]
\begin{minipage}[t]{0.48\linewidth}
 \centering
  \includegraphics[width=8cm,angle=0]{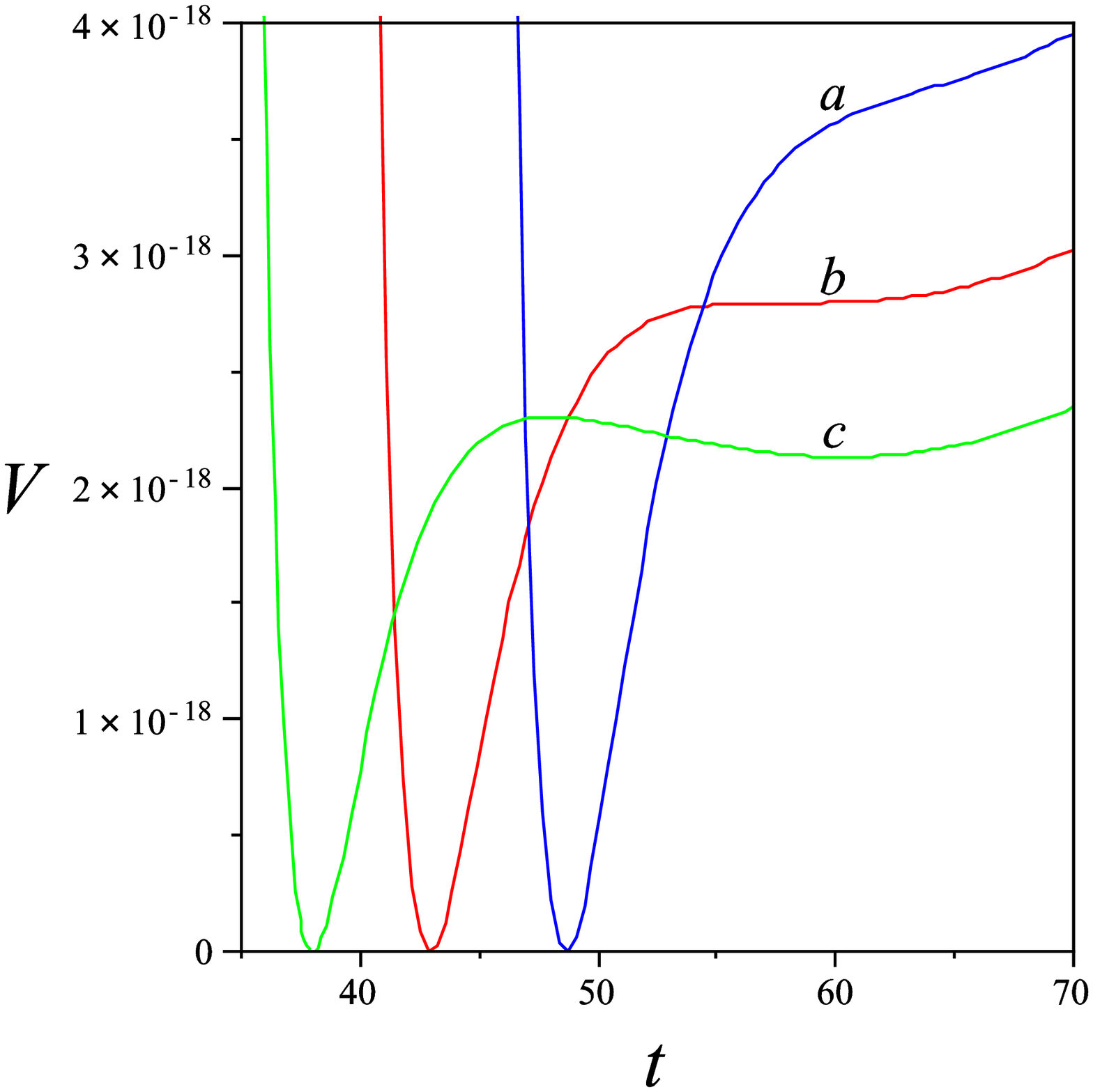}
  \caption{The potential for $\tau=0$ in model
\ref{posneg}. Different lines correspond to different
values of the parameter $C$ with other parameters as in (\ref{par1})
(except $A$ which is always adjusted to ensure the existence of
the SUSY Minkowski minimum):
(a) $C=0.4$, (b) $C = 0.055397$, (c) $C=0.01$.}
  \label{tinf1_C}
\end{minipage}
\hfill
\begin{minipage}[t]{0.48\linewidth}
\centering
  \includegraphics[width=8cm,height=9cm,angle=0]{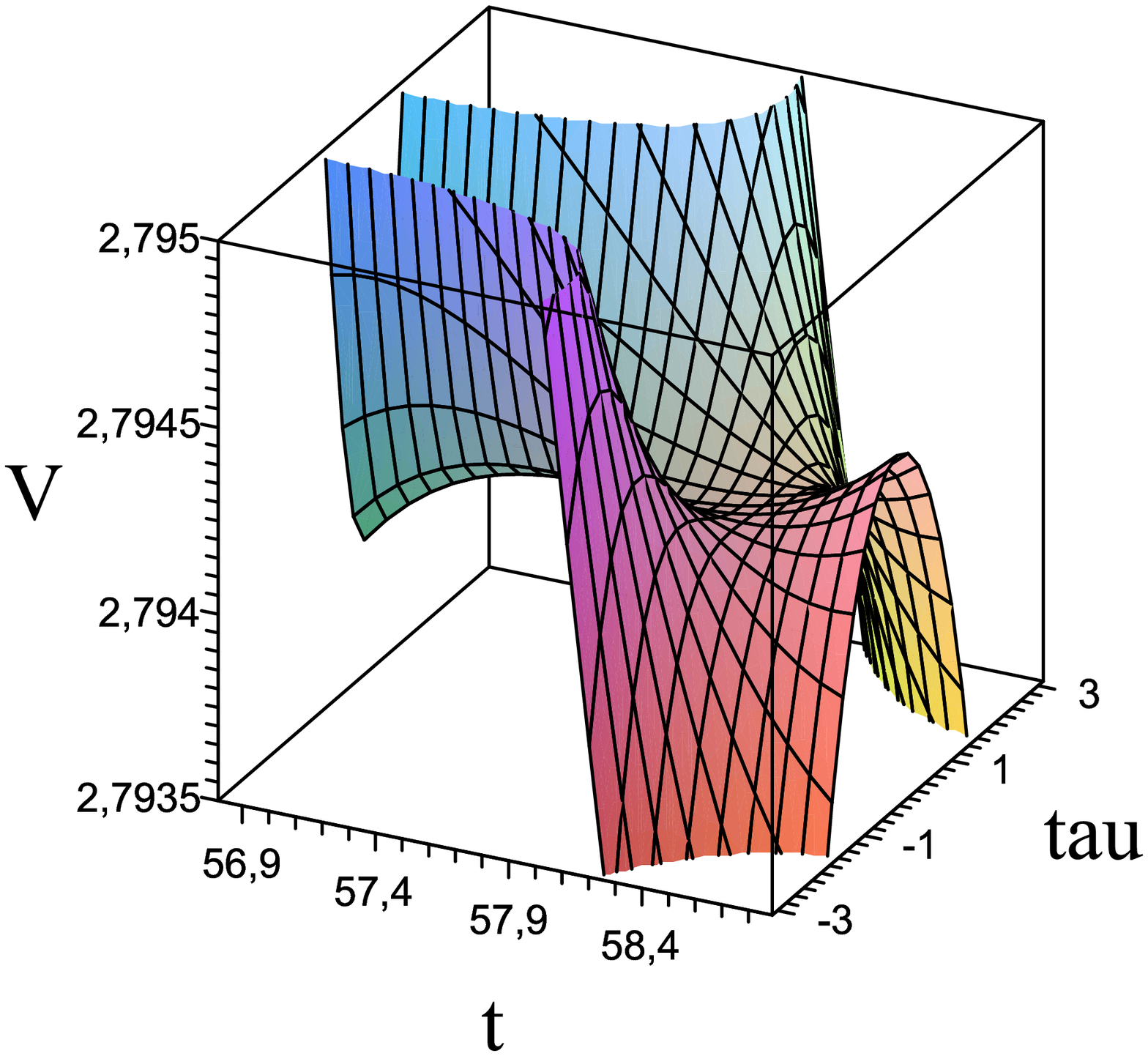}
  \caption{The potential multiplied by $10^{18}$ in the vicinity
of the inflationary inflection point for model \ref{posneg}.}
\label{plot3d1}
\end{minipage}
\end{figure}

In this section we analyze a model with one positive and one negative
exponent in the superpotential (\ref{sup}) and with the K\"ahler
potential with the corrections (\ref{Kcorr}). Let us choose $c<0$, $d>0$.
The conditions (\ref{cCdD}) for the superpotential parameters read:
\begin{equation}
CD>0
\,,
\qquad\qquad
|cC|>|dD|
\,.
\end{equation}
The potential for this model may have a SUSY Minkowski minimum
at $t=T_{\rm Mink}$ given by (\ref{T_Mink}) and a second non-SUSY
minimum at $t>T_{\rm Mink}$. When we  decrease the ratio $C/D$
(keeping other parameters fixed except $A$ which is always
adjusted to keep $W=0$ in the Minkowski minimum) this minimum
becomes more and more shallow and finally disappears.
Close to that transition the potential becomes very flat and allows
for inflation. This is shown in figure \ref{tinf1_C}. Such inflation
ends at a SUSY (near) Minkowski minimum. As a result, the gravitino
mass may be much smaller than the inflationary Hubble constant.

For the numerical example we choose the following set of parameters:
\begin{eqnarray}
\label{par1}
&
A = -8.47423\cdot10^{-7},\hspace{2.13cm}
C = 0.055397 ,\hspace{2.08cm}
D = 2\cdot 10^{-7}
,
&
\nn[4pt]
&
c = -\frac{2\pi}{20},\hspace{2cm}
d = \frac{2\pi}{200},\hspace{2cm}
\xi_{ \alpha '} = 300,\hspace{2cm}
\xi_{ \rm loop} = 300.
&
\end{eqnarray}
Let us first explain why $|c|$ is much bigger than $d$:
In order to trust the perturbative expansion of the K\"ahler potential,
the ratios ${\xi_{\alpha'}}/{(T+\ov{T})^{3/2}}$ and
${\xi_{\rm loop}}/{(T+\ov{T})^2}$ have to be small.
The amount of the corrections required to make the trace of
the $\eta$ matrix positive at the beginning of inflation
is determined by the position of the inflection
point $t_{\rm inf}$. In the model considered in this subsection
$t_{\rm inf}>t_{\rm Mink}$, and the corrections at the Minkowski minimum
are larger than at the inflection point. One can check that
the distance between the Minkowski minimum and the inflection
point decreases for increasing value of the
ratio\footnote
{
Changing the ratio $|c/d|$ one has also to adjust the parameter
$D$ to ensure flatness of the inflection point and the parameter
$A$ to keep $W=0$ at the Minkowski minimum.
}
$|c/d|$. Thus, corrections to the K\"ahler potential
at the Minkowski minimum are smaller for larger $|c/d|$
(increasing  $|c/d|$ we move away from the singularity $t_{\infty}$,
where the potential is not reliable).

Another feature of our numerical example is a very small value
of $D$ as compared to $C$. The reason is quite simple.
In typical racetrack inflationary models both exponents $c$ and $d$
have small (negative) and similar values, because that way it is
easier to obtain
big enough value of $t$ at the minimum of the potential. In our model,
as argued above, $|c|$ and $d$ should not be similar, so as can be seen
from eq.\ (\ref{T_Mink}), the only way to get $|c|t_{\rm Mink}>1$ is
to have a large value of the ratio $C/D$.
Usually, very large ratios of parameters are unnatural. However,
in our model the prefactor $D$ involves the exponential suppression
which comes from the dilaton (which is assumed to be stabilized by fluxes),
so one should expect $D$ to be much smaller than $C$.

For the set of parameters (\ref{par1}), the potential in the
$t$-direction is shown in figure \ref{tinf1_C} (line (b)).
Inflation takes place in the vicinity of the inflection point
located at $t_{\rm inf}\approx57.85$. The potential in the vicinity
of that point is extremely flat, the slow roll parameter
$\epsilon$ is very small: $\epsilon\approx7\cdot10^{-13}$.
The only non zero entry of the $\eta$ matrix at the inflection point is
$\eta_\tau^\tau\approx 1.01$.
The $\tau\tau$ component of the $\eta$ matrix is positive for all $t$
between $t_{\rm Mink}$ and $t_{\rm inf}$, so the $\tau$ field may be
fixed at $0$ during the whole period of inflation.
Starting from the inflection point, inflaton $t$ slowly rolls down
towards the SUSY (near) Minkowski minimum at $t_{\rm mink}\approx42.93$.
The potential has a singularity at $t_{\infty}\approx29.5$.
The ratios ${\xi_{\alpha'}}/{(T+\ov{T})^{3/2}}$
and ${\xi_{\rm loop}}/{(T+\ov{T})^2}$ have the following values:
for the inflationary inflection point around $0.24$ and $0.02$,
respectively, while for the Minkowski minimum around $0.38$
and $0.04$. Therefore, the inflationary inflection point,
as well as the SUSY Minkowski minimum are still in the region of the
potential where the perturbative corrections to the K\"ahler potential
are reasonably small.

In order to find the predictions of our model we have to solve
numerically equations of motion for the fields:
\begin{eqnarray}
\label{eqmt}
t''\br=\br-\left[3-\frac{g_{tt}}{2}\left(t'^2+\tau'^2\right)\right]
\left(g^{tt}\frac{V_t}{V}+t'\right)-\frac{g^{tt}g_{ttt}}{2}
\left(t'^2-\tau'^2\right) \ ,
\\[4pt]
\label{eqmtau}
\tau''\br=\br-\left[3-\frac{g_{tt}}{2}\left(t'^2+\tau'^2\right)\right]
\left(g^{tt}\frac{V_{\tau}}{V}+\tau'\right)-g^{tt}g_{ttt}t'\tau' \ ,
\end{eqnarray}
where $g_{tt}=(g^{tt})^{-1}=\frac{1}{2}\frac{\pa^2 K}{\pa t^2}$,
$g_{ttt}=\frac{1}{2}\frac{\pa^3 K}{\pa t^3}$. We use the number of
e-folds $N\equiv\ln a(t_{\rm cosm})$ instead of the cosmic time
$t_{\rm cosm}$ as the independent variable.
In the above equations $'$ denotes derivatives with respect to $N$.
We solve numerically these equations with the initial conditions:
$\tau(0)=0$, $t'(0)=\tau'(0)=0$ and $t(0)$ at or close to the
inflection point $t_{\rm inf}=57.85$.
We find the constant value of $\tau=0$ during the whole period of
inflation, as was anticipated earlier. For $t(0)=t_{\rm inf}$,
after $195$ e-folds of inflation, $t$ starts to oscillate in the
vicinity of the SUSY (near) Minkowski minimum, where inflation ends.

\begin{table}[t]
\centering
\begin{tabular}{|c|c|c|c|c|c|c|c|c|c|c|c|c|}
\hline	
$\Delta t_{\rm ini}$ & $-0.1$ & $-0.02$ & $-0.01$ & $-0.001$ & $0$
& $0.001$ & $0.01$ & $0.1$ & $0.2$ & $0.3$ &  $0.4$ & $0.5$
\\ \hline
$N$ & $12$ & $53$  & $90$  & $181$  &  $195$
& $209$ &  $300$  & $375$ &  $380$  &  $381$ &  $382$   & $382$
\\ \hline
\end{tabular}
\caption{Number of e-folds of inflation for several values of
$\Delta t_{\rm ini}= t_{\rm ini}-t_{\rm inf}$ in model
\ref{posneg}.}
\label{tab}
\end{table}

We can now ask: how much fine-tuned should be the initial conditions
for the inflaton? In table \ref{tab} we show how the duration of
inflation depends on the initial value of the inflaton.
About 380 e-folds are obtained for a relatively wide range of
$t_{\rm ini}$ between $(t_{\rm inf}+0.1)$ and $(t_{\rm inf}+0.5)$.
The fine-tuning of the initial conditions (defined as
$\left|\Delta t_{\rm ini}\right|/t_{\rm inf}$) necessary to obtain more
than $60$ e-folds of inflation is at the level of one percent
for positive $\Delta t_{\rm ini}$ and about one permille for negative
$\Delta t_{\rm ini}$.

Taking into account only the flatness of the potential in the
$t$-direction one could expect that even less fine-tuning is necessary.
By solving equation of motion for the inflaton $t$ (\ref{eqmt})
with $\tau=\tau'=0$ one can check that $60$ e-folds of inflation
is obtained for $\Delta t_{\rm ini}\lesssim3.3$.
This corresponds to $\left|\Delta t_{\rm ini}\right|/t_{\rm inf}\approx0.06$.
However, solving (\ref{eqmt}) independently of (\ref{eqmtau}) is justified
only in the region where $\left.\left(\pa^2 V/\pa\tau^2\right)\right|_{\tau=0}$
is positive. We find that for $\Delta t_{\rm ini}\gtrsim0.5$ the field
$\tau$ becomes tachyonic (as seen in figure \ref{plot3d1}),
the trajectory $\tau=0$ becomes unstable and
the fields do not evolve towards the Minkowski minimum.
The value of $\Delta t_{\rm ini}$, for which the mass of the field $\tau$
is positive, can be enlarged by increasing the corrections to the
K\"ahler potential. However, it is not desirable since we want to
keep these corrections as small as possible. Nevertheless, the
fine-tuning of the initial conditions in the inflection point
inflation is much smaller than in the case of the saddle point inflation.

Finally, we comment on the fine-tuning of the parameters.
We need two fine-tunings: one to obtain a light gravitino
(the SUSY minimum must be near Minkowski) and second to
have, for some region of $t$, a very flat potential suitable for
inflation. In both cases one has to tune appropriate combinations
of all parameters. For simplicity, we tune two parameters, $A$ and $C$,
keeping other parameters fixed at some ``round'' values.
The fine-tuning of $A$ must be at the level of $10^{-5}$ in order
to have a TeV-range gravitino mass (this fine-tuning can be relaxed if
one wants the gravitino to be heavier). The parameter $C$
must be tuned at the level of $10^{-4}$ in order to construct
flat enough inflection point to obtain at least $60$ e-folds of inflation.
Notice that the fine-tuning in this model is much weaker than in
the triple gaugino condensation model \cite{bo}, which was the first
model of moduli inflation with a TeV-range gravitino mass. We recall that
in the triple gaugino condensation model, flatness of the potential was
obtained by fine-tuning of two parameters at the level of $10^{-5}$ and
$10^{-7}$. In that model two fine-tunings were needed because the diagonal
and off-diagonal entries of the $\eta$ matrix had to be tuned
separately.
The reason is that in models with a SUSY Minkowski minimum the trace
of the $\eta$ matrix cannot be large, because it is made positive only
due to subleading corrections which are supposed to be small. Therefore,
it is not possible to tune the smallest eigenvalue of the $\eta$ matrix
using only one parameter, unless off-diagonal entry is suppressed by
some mechanism.
In all models presented in this paper, inflation takes place in the
region where $\tau=0$ so the off-diagonal entry of the $\eta$ matrix
vanishes automatically (when the parameters of the superpotential
are real). Therefore, that additional fine-tuning found in \cite{bo}
is not required.


\subsection{Two positive exponents}
\label{twopos}

Now we consider models with the superpotential (\ref{sup}) with
both exponents positive. Without loosing generality, we consider $c>d>0$.
In such case the conditions (\ref{cCdD}) for the superpotential
parameters read:
\begin{equation}
CD<0
\,,
\qquad\qquad
c|C|<d|D|
\,.
\end{equation}
Inflation, that ends in a SUSY (near) Minkowski minimum,
can be realized in this model if the K\"ahler potential (\ref{Kcorr})
with the leading string corrections is used. The mechanism of \
inflation is very similar to the one presented in the previous
subsection. For some region of the parameter space, a non-SUSY
minimum with a positive cosmological constant may exist. By changing
the parameter $C$ one can obtain a very flat inflection point.
However, there is one important difference between this model and the
previous one. Namely, in this model $t_{\rm inf}<t_{\rm Mink}$, so the
corrections to the K\"ahler potential at the SUSY (near) Minkowski
minimum are always smaller than at the inflationary inflection point.
Therefore, one can keep the SUSY (near) Minkowski, as well as the
inflationary inflection point, far away from the singularity of the
potential. This also implies that the inflationary model can be
constructed for arbitrary values of parameters $c$ and $d$,
which is the main advantage of this model as compared to the one
with only one positive exponent. For the numerical example we
choose the following set of parameters:
\begin{eqnarray}
&
A = 2.17351\cdot10^{-6},\hspace{2cm}
C = 1.273737\cdot10^{-8} ,\hspace{2cm}
D = 4\cdot10^{-8} ,
&
\nn[4pt]
&
c = \frac{2\pi}{60},\hspace{2.25cm}
d = \frac{2\pi}{70},\hspace{2.35cm}
\xi_{ \alpha '} = 200,\hspace{2.4cm}
\xi_{ \rm loop} = 200.
&
\label{par2}
\end{eqnarray}
\begin{figure}
\begin{minipage}[t]{0.48\linewidth}
 \centering
  \includegraphics[width=8cm,angle=0]{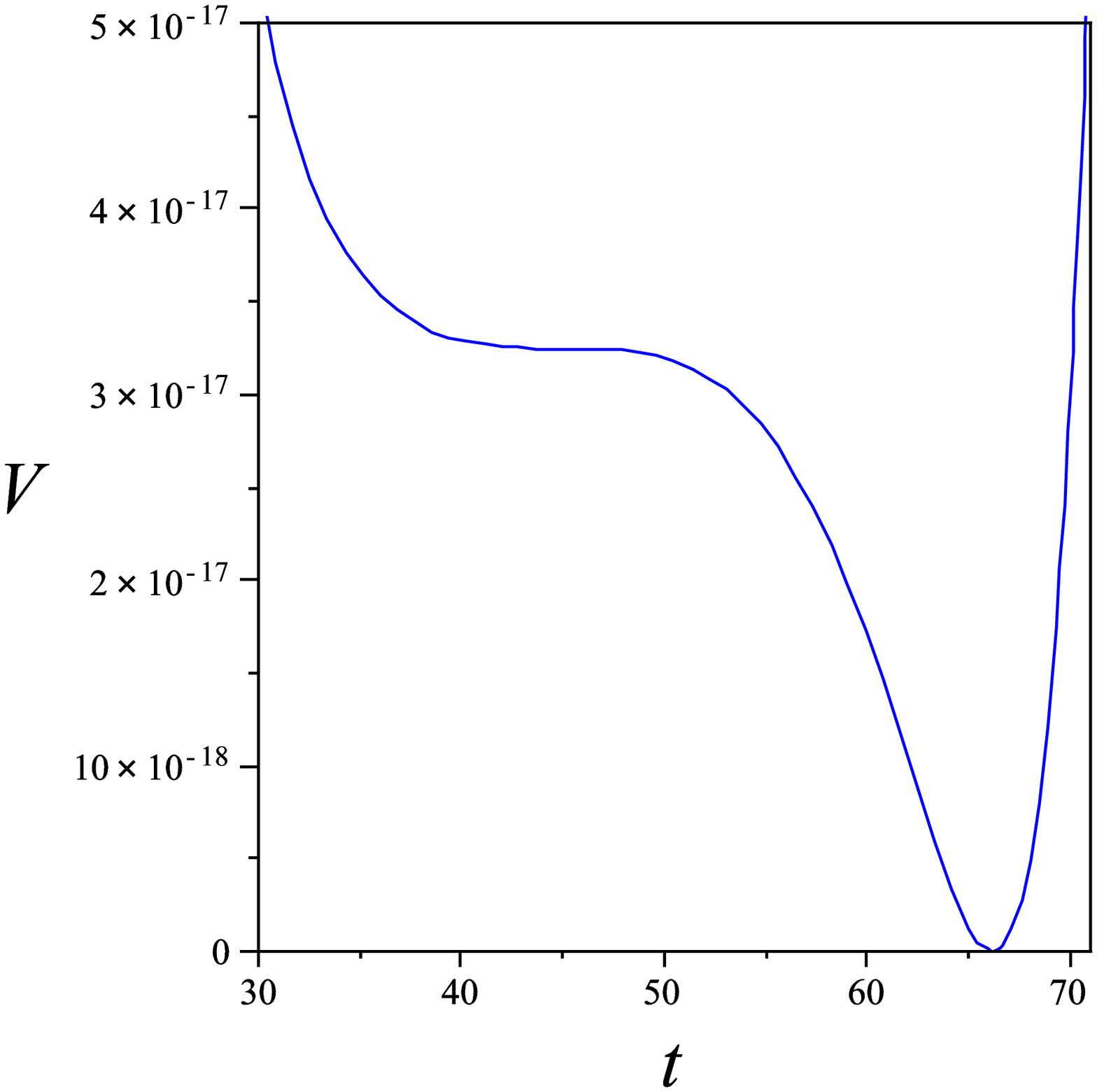}
  \caption{The potential for $\tau=0$ in model
\ref{twopos} with the parameters as in (\ref{par2}).}
  \label{plott2}
\end{minipage}
\hfill
\begin{minipage}[t]{0.48\linewidth}
  \centering
  \includegraphics[width=8cm,height=9cm,angle=0]{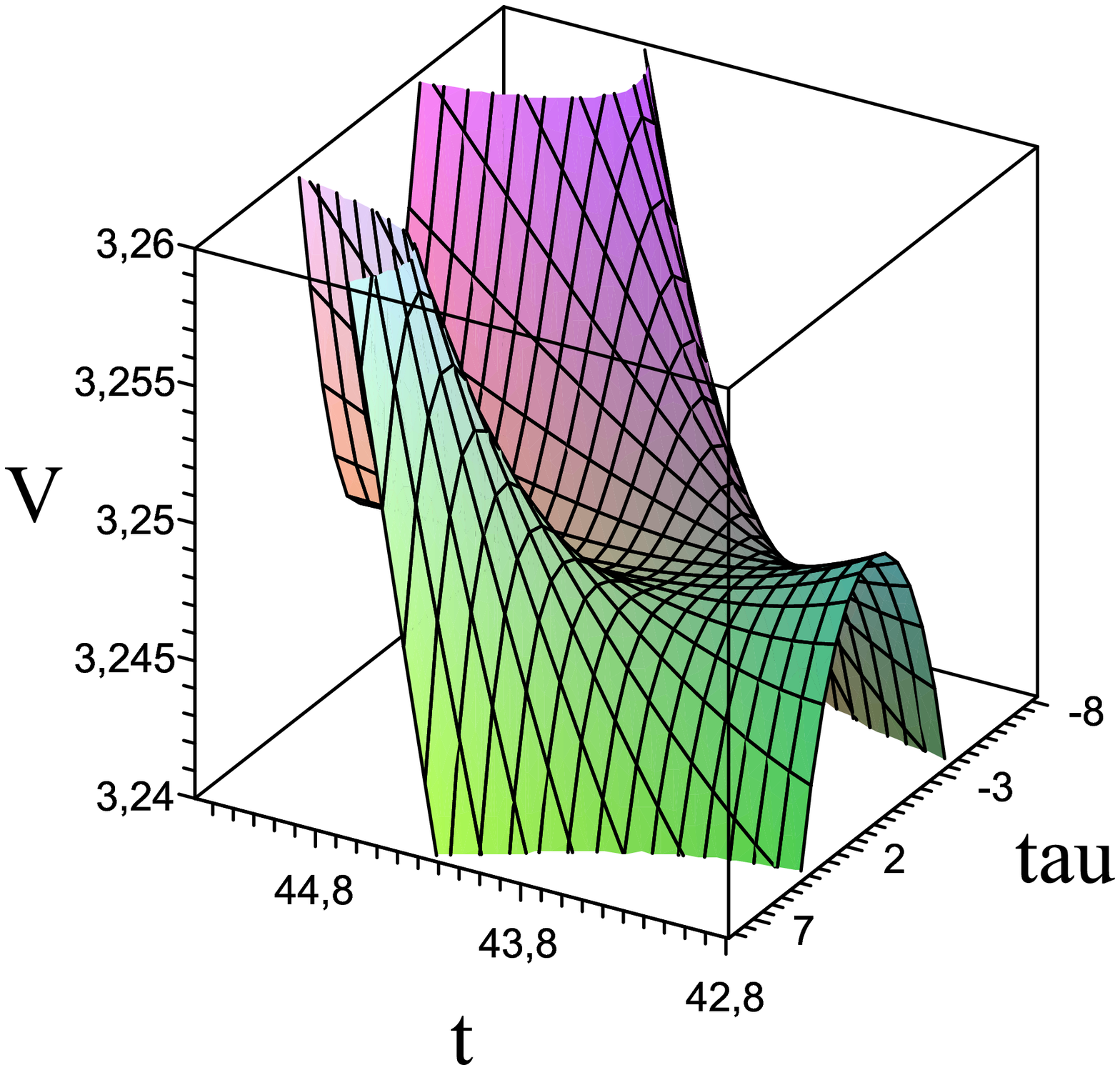}
  \caption{The potential multiplied by $10^{17}$ in
the vicinity of the inflationary inflection point
for model \ref{twopos}.}
\label{plot3d2}
\end{minipage}
\end{figure}

The potential in the $t$-direction is shown in figure \ref{plott2}.
Inflationary inflection point occurs at $t_{\rm inf}\approx44.84$, where
the $\eta$ matrix has the following entries:
$\eta_t^t=\eta_\tau^t=0$, $\eta_\tau^\tau\approx 1.21$.
The parameter $\epsilon\approx6\cdot10^{-13}$ is small enough to
allow for a long period of inflation. The SUSY Minkowski
minimum is situated at $t_{\rm Mink}\approx66.19$.

Since in this model $t_{\rm inf}>t_{\rm Mink}$, the corrections in the
region important for inflation are smaller than in a model
\ref{posneg} with one positive and one negative exponent. We find the ratios
${\xi_{\alpha'}}/{(T+\ov{T})^{3/2}}$ and
${\xi_{\rm loop}}/{(T+\ov{T})^2}$ at the inflationary inflection point
equal to $0.24$ and $0.02$, respectively, while at the Minkowski minimum
$0.13$ and $0.01$. The singularity of the potential is situated at
$t_{\infty}\approx22.86$. Therefore, inflation takes place in the region of
the potential, where the perturbative expansion of the K\"ahler
potential is justified.

In contrast to the model \ref{posneg} with one positive and one
negative exponent, in the present model with both exponents positive
there is no hierarchy between the parameters $C$ and $D$. However, both
of them are very small. Similarly to the previous model, the smallness of
these parameters can be explained by the exponential suppression coming
from a large vev of the dilaton.

Using equations of motion (\ref{eqmt})-(\ref{eqmtau}) we evolved the
fields starting from the inflection point. The evolution of the inflaton
$t$ ends in the Minkowski minimum after $277$ e-folds of inflation.
Similarly as in the previous model, it turns out that the inflation
can be longer if it starts at some region above the inflection
point. This is demonstrated in table \ref{tab2}. Inflation can be as
long as about $540$ e-folds. The fine-tuning of the initial conditions
is at the level of a few percent:
$-3\cdot10^{-2}\lesssim\Delta t_{\rm ini}/t_{\rm inf}\lesssim6\cdot10^{-4}$.
The upper bound on $\Delta t_{\rm ini}$ is not due to the potential shape
in the $t$
direction\footnote
{
By solving the equation of motion (\ref{eqmt}) for the inflaton $t$
with $\tau$ fixed at zero,
one would obtain $60$ e-folds of inflation even
for $\Delta t_{\rm ini}\sim-4$, corresponding to
$\left|\Delta t_{\rm ini}\right|/t_{\rm inf}\approx0.09$.
}
but from the requirement of stability in the $\tau$
direction as can be seen from figure \ref{plot3d2}.

\begin{table}[t]
\centering
\begin{tabular}{|c|c|c|c|c|c|c|c|c|c|c|c|c|}
\hline	
$\Delta t_{\rm ini}$ & $-1.2$ & $-1$ &  $-0.5$ & $-0.3$ & $-0.1$
& $-0.01$ & $-0.001$  & $0$ & $0.001$ & $0.01$ & $0.03$ & $0.1$
\\
\hline
$N$ & $546$ & $547$ &  $546$  &  $544$  & $535$
& $429$ & $298$  &  $277$  & $256$  & $124$ & $50$  & $16$
\\
\hline
\end{tabular}
\caption{Number of e-folds of inflation for several values of
$\Delta t_{\rm ini}= t_{\rm ini}-t_{\rm inf}$ in model
\ref{twopos}.
}
\label{tab2}
\end{table}

The fine-tuning of the parameters is comparable to the one in model
\ref{posneg} with one positive and one negative exponent.
The parameter $A$ has to be fine-tuned at
the level of $10^{-5}$ to obtain gravitino mass of the order of TeV.
The flatness of the potential is achieved due to the fine-tuning of the
parameter $C$ at the level of $10^{-5}$.


\subsection{Two negative exponents}
\label{KL}

A model with the superpotential (\ref{sup}) with two negative exponents
is nothing else as the KL model \cite{KaLi}. It was shown in \cite{bo}
that in such model it is not possible to realize inflation to a (near)
Minkowski minimum from a saddle point. It is also not possible to have
such inflation starting close to an inflection point. The reason is very
simple: The second minimum in the $t$-direction (existing in addition
to the SUSY Minkowski one) is supersymmetric and has negative value
of the potential. It can not be deformed to an inflection point with
positive energy by any changes of the parameters if both exponents
$c$ and $d$ remain negative. Thus, to realize inflation in the KL
model one has to add some uplifting potential. One uplifts the SUSY AdS
minimum to a (near) Minkowski one and the second minimum to an
inflection (or saddle) point. Such models
were investigated for example in \cite{LiWe}. Unfortunately, in models
with uplifting the gravitino mass is bigger than the Hubble constant
during inflation and the clash between light gravitino and high scale
inflation appears.
Nevertheless, we investigate in some detail also such model in order
to compare its features, for example fine-tuning of the
parameters and initial conditions, with other models with two
gaugino condensates.

Following \cite{LiWe}, we add to the potential (\ref{potdouble}) the
following uplifting term:
\begin{equation}
\label{uplift}
\Delta V=\frac{E}{t^2} \ .
\end{equation}
The uplifting gives positive contribution to the trace of the
$\eta$ matrix and it is not necessary to add any corrections to
the K\"ahler potential. Moreover, corrections of the form (\ref{Kcorr}),
when taken into account, change the results only slightly, so in
our numerical calculations we use the leading order K\"ahler
potential (\ref{leadK}).

The uplifting term (\ref{uplift}) does not significantly change
the positions of the stationary points of the potential.
Therefore, the position of the inflationary inflection point can
be well approximated by that of the SUSY Minkowski minimum (\ref{T_Mink}):
\begin{equation}
t_{\rm inf}\approx\frac{1}{d-c}\ln\left({-\frac{cC}{dD}}\right) \,.
\end{equation}
Without loosing generality, we consider $c<d<0$. Therefore, the
parameters in this model should satisfy the conditions (\ref{cCdD}):
\begin{equation}
CD<0
\,,
\qquad\qquad
c|C|<d|D|
\,.
\end{equation}
Two tunings of the parameters are necessary: one to get (almost) vanishing
vacuum energy and another to obtain a sufficiently flat inflection point.
Technically we choose to tune $A$ and $E$ with other parameters fixed.
However, it turns out that those other parameters also play an important
role. They influence the height of the barrier separating the Minkowski
minimum from the run-away region at large values
of the volume $t$. There are regions in the parameter space for which
inflation lasts long enough but at the end the inflaton runs away to
infinity instead of oscillating around the vacuum at finite $t$.
We found that the barrier height grows with increasing $|C/D|$ and/or
decreasing $|c/d|$ when the value of the potential at the inflection point
is kept fixed (in order to stay in agreement with COBE normalization).

\begin{figure}
\begin{minipage}[t]{0.48\linewidth}
\centering
  \includegraphics[width=8cm,angle=0]{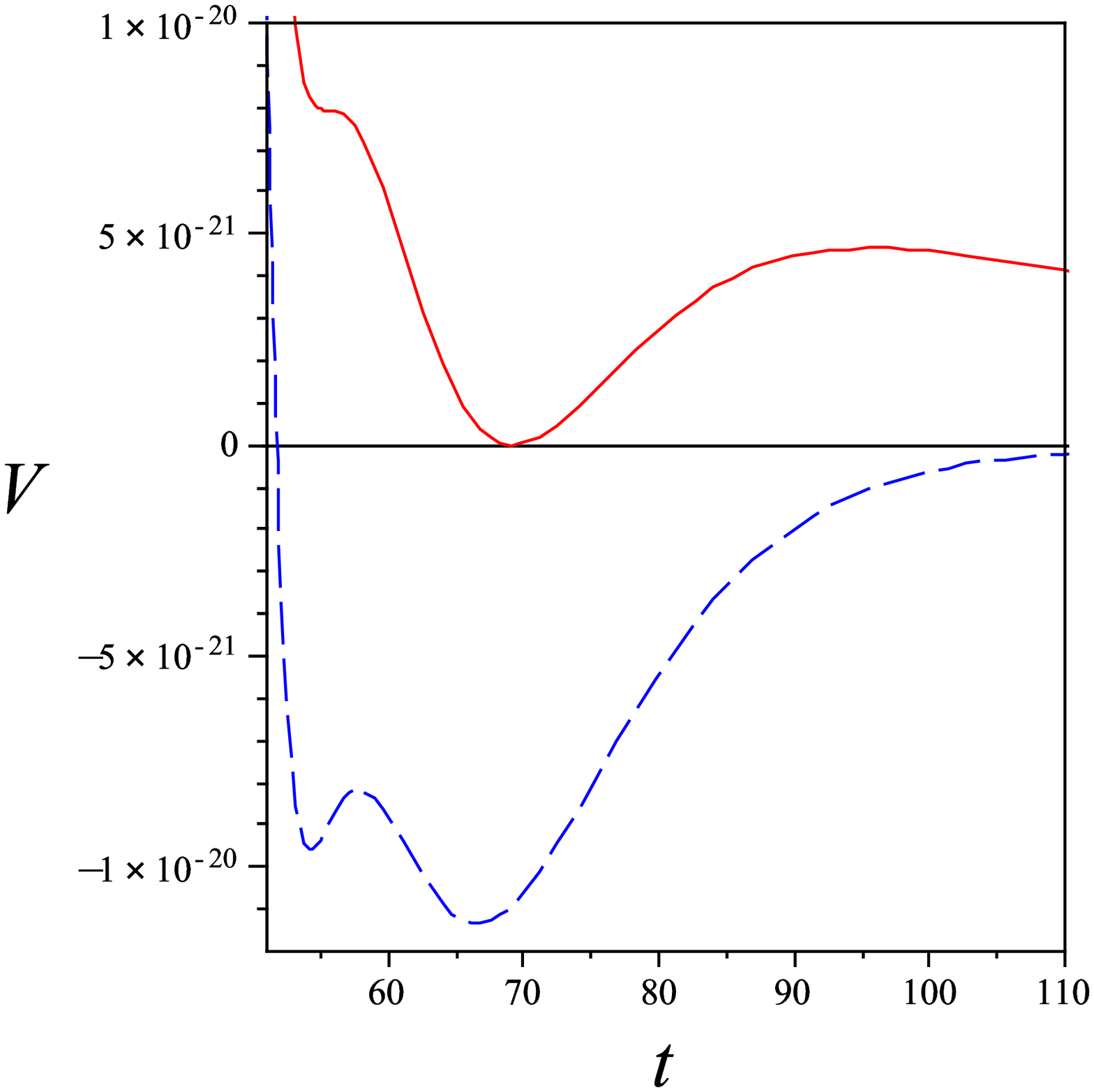}
  \caption{The potential for model \ref{KL}
with the parameters (\ref{parKL}) with (solid line)
and without (dashed line) uplifting.}
\label{plottKL}
\end{minipage}
\hfill
\begin{minipage}[t]{0.48\linewidth}
  \centering
  \includegraphics[width=8cm,height=8.5cm,angle=0]{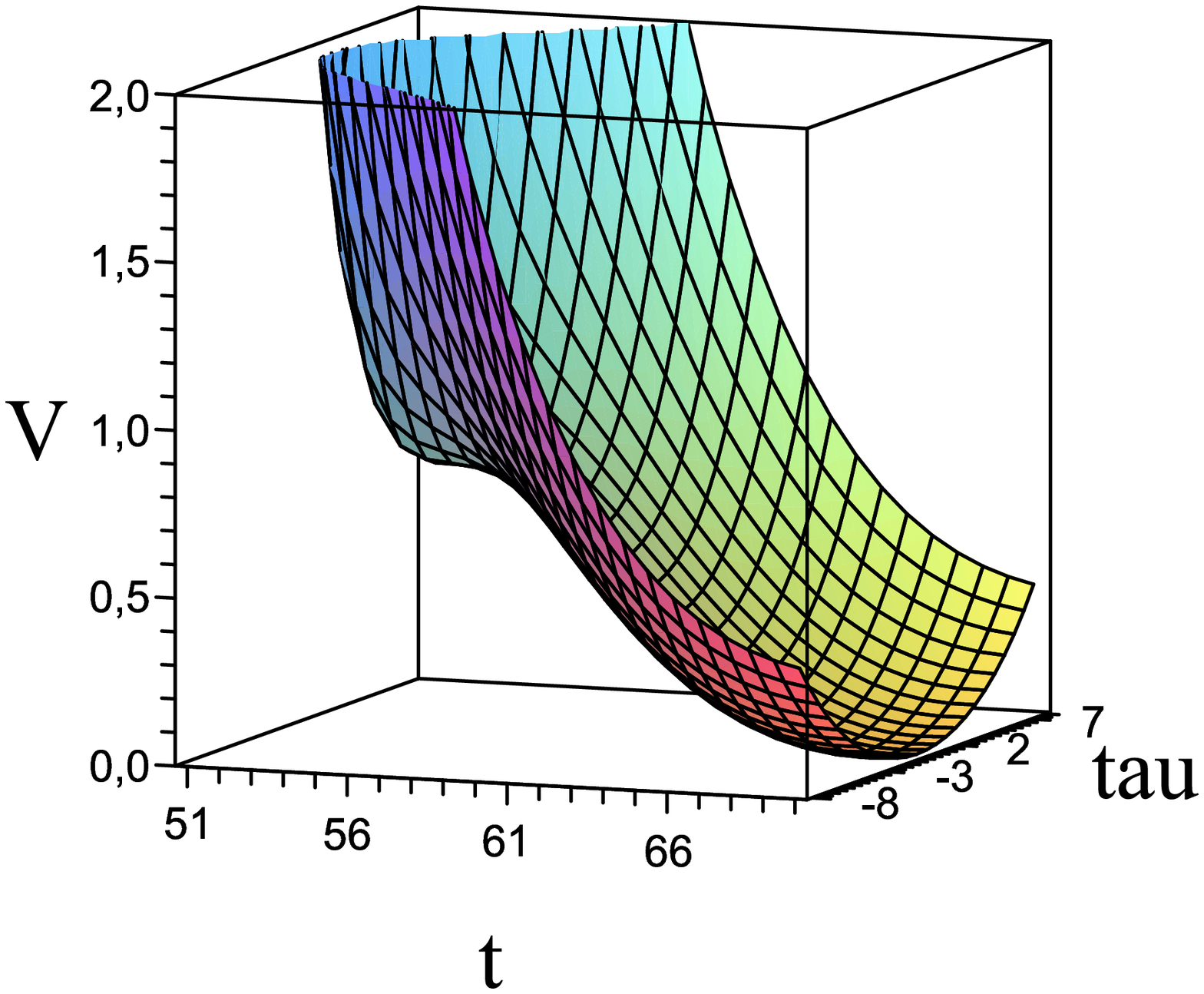}
  \caption{The potential multiplied by $10^{20}$ in
the vicinity of the inflationary inflection point
for model \ref{KL}.}
\label{plot3dKL}
\end{minipage}
\end{figure}

For the numerical example we choose the following set of parameters:
\begin{equation}
\begin{tabular}{lll}
$A=1.20533994\cdot10^{-7} , \qquad$ &
$C=4\cdot10^{-4} \,, \qquad$ &
$D=-4\cdot10^{-5} \,,$
\\[4pt]
$E=5.2220577\cdot10^{-17} \, , \qquad$ &
$c=-\frac{2\pi}{40} \, , \qquad$ &
$d=-\frac{2\pi}{60} \, .$
\end{tabular}
\label{parKL}
\end{equation}
The potential in the $t$-direction before and after uplifting is shown in
figure \ref{plottKL}. The inflationary inflection point is generated by
uplifting of the minimum at smaller volume and is situated at
$t_{\rm inf}\approx55.59$.
At this point, the only non-vanishing entry of the $\eta$
matrix is $\eta^{\tau}_{\tau}\approx191$. Notice that the trace of the $\eta$
matrix is much larger than in models \ref{posneg} and \ref{twopos}
with the SUSY (near) Minkowski
minima. The reason is that a large positive contribution to the trace of
the $\eta$ matrix is generated by a large uplifting term. At the same time,
such large uplifting generates a large gravitino mass. In our example the
gravitino mass at the vacuum is about $10^8$ GeV which is several orders of
magnitude more than in typical models with low scale SUSY breaking.
The Minkowski vacuum occurs at $t_{\rm  vac}\approx69.2$.
We find a very small value of $\epsilon\approx3\cdot10^{-15}$ at the
inflection point. In this model
$\tau=0$ is a minimum in the $\tau$-direction for an arbitrary
value of $t$, as can be seen in figure \ref{plot3dKL}.

\begin{figure}[t!]
  \centering
  \includegraphics[width=10cm,angle=0]{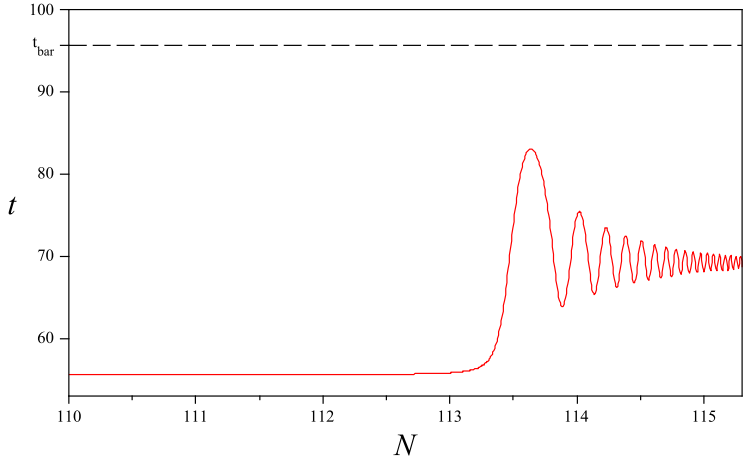}
  \caption{Evolution of the inflaton $t$ in the last stage of inflation
as a function of e-folds $N$ in model \ref{KL}. The dashed line shows
the position of the barrier separating the vacuum at finite $t$ from the
run-away region.}
  \label{tevolKL}
\end{figure}
\begin{table}
\centering
\begin{tabular}{|c|c|c|c|c|c|c|c|c|c|c|}
\hline	
$\Delta t_{\rm ini}$ & $-0.05$ & $-0.04$ &  $-0.03$ & $-0.01$
& $-0.001$ & $-0.0001$ & $0$ & $0.0001$ & $0.0002$ & $0.001$
\\
\hline
$N$ & $11$  & $209$ &  $222$  &  $224$
& $211$ &  $144$ &  $113$ & $85$ & $64$  & $19$
\\
\hline
\end{tabular}
\caption{
Number of e-folds of inflation for several values of
$\Delta t_{\rm ini}= t_{\rm ini}-t_{\rm inf}$ in model
\ref{KL}.
}
\label{tabKL}
\end{table}

Solving numerically the equations of motions (\ref{eqmt})-(\ref{eqmtau}) with
the initial conditions exactly at the inflection point $t(0)=t_{\rm inf}$, we
obtain about $113$ e-folds of inflation.
Starting from some region above the inflection point one can
obtain longer period of inflation, in some cases exceeding $220$ e-folds
(see table \ref{tabKL}). It can be seen from figure \ref{plottKL} that
the value of the potential in the vicinity of the inflection point
is bigger than the height of the barrier separating the vacuum from
the run-away region. Nevertheless, inflation ends with the inflaton
oscillating around the vacuum (see figure \ref{tevolKL}).
The overshooting does not occur due to the Hubble friction.
Our numerical example was chosen in such a way that the height of the barrier
is close to the minimal one necessary for protecting the inflaton from
running away to infinity. The reason is that with such choice we minimize
the necessary fine-tuning of parameters. We explain this point below.

In the presented example the parameter $A$ has to be fine-tuned at the level
of $10^{-8}$ to provide more than $60$ e-folds of inflation. Moreover, in
contrast to models with the SUSY (near) Minkowski minimum, how much
fine-tuning of the parameter $A$ is required depends quite strongly on other
parameters of the model. We found that the fine-tuning of $A$ becomes stronger
when one decreases the ratio $|c/d|$ and/or increases the ratio $|C/D|$. As we
mentioned before, these two ratios control also the height of the barrier that
protects the Minkowski vacuum from overshooting. Therefore, the fine-tuning
of the parameter $A$ (responsible for the flatness of the potential) is
related to the height of this barrier. More fine-tuning is required when
the barrier is higher. The fine-tuning at the level of $10^{-8}$ is
the minimal one which allows the inflaton to finish its evolution in
the vacuum at finite $t$. The height of the barrier influences in a
similar way also the fine-tuning of the initial conditions: the
fine-tuning becomes stronger for higher barrier. Unfortunately, even for
the lowest possible barrier the fine-tuning is much stronger than in
models with SUSY (near) Minkowski minima. For our numerical example
in order to obtain at least 60 e-folds of inflation the initial
conditions must satisfy
$-8\cdot10^{-4}\lesssim\Delta t_{\rm ini}/t_{\rm inf}\lesssim4\cdot10^{-6}$.

The relation between the fine-tuning (of parameters and initial
conditions) and the height of the barrier separating the vacuum
from the run-away region can be explained by the following reasoning.
The size of the region $\left|\Delta t_{\rm ini}\right|/t_{\rm inf}$
from which inflation can start and last more than $60$ e-folds,
depends on the distance (before uplifting) between the minimum and the
maximum that generate inflection point after uplifting of the potential.
If we insist on the existence of a high barrier, the value of the
potential before uplifting at the maximum should be much smaller than
zero. This follows from the fact that the uplifting term (\ref{uplift})
lifts much more strongly this maximum than the region of the potential
from which the barrier arises. On the other hand,
lowering the maximum (before uplifting) implies that it becomes closer
to the minimum from which inflection point is generated after uplifting.
Therefore, the size of the region
$\left|\Delta t_{\rm ini}\right|/t_{\rm inf}$ from which inflation can
start becomes smaller. Notice that for smaller region of the flat potential
the slow-roll parameter $\epsilon$ (which measures the flatness of the
potential) has to be smaller than in the case of larger
$\left|\Delta t_{\rm ini}\right|/t_{\rm inf}$ in order to obtain the same
number of e-folds. This is because the inflaton has to move slower if the
region where slow-roll conditions are satisfied is smaller. Thus,
for smaller values of $\left|\Delta t_{\rm ini}\right|/t_{\rm inf}$
the parameter $A$ (which is responsible for small value of $\epsilon$) has
to be more fine-tuned.
In the case when the value of the potential at the maximum (before
uplifting) is positive, the distance
to the minimum is larger and less fine-tuning of the parameter $A$ is needed.
However, the price of this smaller fine-tuning is a very low barrier that
protects the inflaton from runaway to infinity.
Finally, we point out that the above reasoning does not depend on a
specific form of the potential. Therefore, it applies to all other models
where uplifting generates simultaneously inflationary inflection point
and the Minkowski minimum (e.g.\ to the model discussed later
in subsection \ref{negthr}).


\section{Models with single gaugino condensation and threshold corrections}
\label{1condensatethr}

In this section we discuss models with only one gaugino condensate.
It is known that superpotential (\ref{sup}) with $D$ set to zero
is not suitable for describing inflation even with nonstandard K\"ahler
potential. Therefore, we assume that the threshold corrections introduce
some field dependence of the prefactor of the gaugino condensation term.
The simplest such superpotential reads
\begin{equation}
W=A+\left(C_0+C_1T\right)e^{cT}
\,.
\label{Wthr}
\end{equation}
This kind of superpotential with $C_0=0$ was considered for
example in \cite{Krefl} but in that case inflation does not
work as will be explained later. For $C_0\ne0$ the superpotential
(\ref{Wthr}) depends only on 4 parameters.
This is one parameter less than in models with two
gaugino condensates considered in section \ref{2condensates}.
Nevertheless, also in this case a SUSY Minkowski minimum exists
due to a $T$-dependent prefactor if the superpotential parameters
satisfy the following condition
\begin{equation}
A=\frac{C_1}{c}\exp\left(-\frac{cC_0}{C_1}-1\right) \ .
\label{Athr}
\end{equation}
The position of that minimum is given by
\begin{equation}
T_{\rm Mink}=-\frac{1}{c}-\frac{C_0}{C_1} \ ,
\label{minkthr}
\end{equation}
and is real for real parameters in (\ref{Wthr}). Similarly to the case
considered in the previous section, potential with the corrections to
the K\"ahler potential is very complicated. Therefore, we present
only the leading part of the potential for the modulus $t$:
\begin{equation}
V=\frac{e^{ct}}{6t^2}{ \left[ C_{{0}}c+C_{{1}}\left(ct+1\right)\right]
 \left[\left(C_{{0}}\left(ct-3\right)+C_{{1}}t\left(ct-2\right)   \right)
{e^{ct}}-3A \right] }+\ldots \,,
 \label{Vthr}
\end{equation}
where we set $\tau=0$ because the axion may be fixed to zero
during and after inflation. The ellipsis stands for subleading terms
involving $\xi_{\alpha'}$ and $\xi_{\rm loop}$, which do not change
qualitatively the features of the potential in the $t$ direction.
In the following subsections we will investigate models with two possible
signs of the exponent.


\subsection{Positive exponent}
\label{posthr}

Let us start with the model with positive exponent, $c>0$,
in the superpotential (\ref{Wthr}).
In order to ensure that the volume at the SUSY Minkowski
minimum (\ref{minkthr}) is positive, $t_{\rm Mink}>0$,
the parameters have to satisfy $(-cC_0/C_1)>1$
(so, for positive $c$, different signs of $C_0$ and $C_1$ are required).
The parameter $A$ is fixed by the condition
(\ref{Athr}) to ensure existence of a  SUSY Minkowski minimum. One
can show that the slow-roll parameters depend on $c$,
$C_0$ and $C_1$ only through the combination $(cC_0/C_1)$.
So, this combination is effectively the only free parameter in the
class of models considered in this subsection. However, even though
there is so little freedom in the parameter space, slow-roll inflation
can be realized. Again, the key property of the potential, that allows
for inflection point inflation ending in a SUSY (near) Minkowski minimum,
is the existence of a non-SUSY minimum for some region of the parameter
space. Such a minimum exists for sufficiently large values of the ratio
$|cC_0/C_1|$. Choosing $|cC_0/C_1|$ slightly smaller than the
critical value for which such minimum disappears, one obtains a very flat
inflection point, where inflation can take place.
To ensure that at this inflection point the mass squared of the field
$\tau$ is positive, one has to turn on the corrections to the K\"ahler
potential (\ref{Kcorr}). For the numerical example we choose the
following set of parameters:
\begin{equation}
\begin{tabular}{lll}
$A=-3.81013\cdot10^{-6} , \qquad$ &
$C_0=5\cdot10^{-8} , \qquad$ &
$C_1=-6.12725\cdot10^{-10}\, ,$
\\[4pt]
$c=\frac{2\pi}{70}\, , \qquad$ &
$\xi_{ \alpha '} = 200\, , \qquad$ &
$\xi_{ \rm loop} = 200\, .$
\end{tabular}
\label{par3}
\end{equation}

The shape of the potential in the $t$-direction is similar
to that for the model \ref{twopos},
shown in figure \ref{plott2}. In the present case the inflationary
inflection point occurs at $t_{\rm inf}\approx47.12$
where the only non-vanishing entry of the $\eta$ matrix is
$\eta_{\tau}^{\tau}\approx0.77$ and the $\epsilon$ parameter is very small:
$\epsilon\approx3\cdot10^{-11}$.
The position of the SUSY (near) Minkowski minimum, where inflation ends,
is found to be $t_{\rm Mink}\approx70.46$, while the singularity of the
potential is situated at $t_{\infty}\approx22.86$. This singularity is
not very close to the region crucial for inflation, so one may expect
that the corrections to the K\"ahler potential are not very large.
To check this we compute the ratios
${\xi_{\alpha'}}/{(T+\ov{T})^{3/2}}$ and
${\xi_{\rm loop}}/{(T+\ov{T})^2}$.
For the inflationary inflection point they are, respectively,
around $0.22$ and $0.02$. Since in this model $t_{\rm Mink}>t_{\rm inf}$,
the corrections at the Minkowski minimum are smaller.
For our numerical example they are around $0.12$ and $0.01$.
Similarly to other models with
SUSY (near) Minkowski minima, we can conclude that the
inflation takes place in the region of the potential where the
subleading corrections are still reasonably small.

Integrating numerically the equations of motion (\ref{eqmt})-(\ref{eqmtau})
with the initial conditions $t(0)=t_{\rm inf}$, $\tau(0)=\tau'(0)=t'(0)=0$,
we obtain inflation which after about $113$ e-folds ends in the
SUSY Minkowski
minimum. Table \ref{tab3} shows how much longer inflation can be if
it starts at some region above the inflection point.
The maximal duration of inflation is of about $220$ e-folds.
The fine-tuning of the initial conditions is at the level of a few
percent ($-0.02\lesssim\Delta t_{\rm ini}/t_{\rm inf}\lesssim5\cdot10^{-4}$).
For $\Delta t_{\rm ini}\lesssim-0.9$ the potential is unstable in the
$\tau$
direction\footnote{
Ignoring such instability in the $\tau$ direction, and
taking into account only the $t$-dependence of the potential, one would
find long enough inflation even for $\Delta t_{\rm ini}\approx-5$
corresponding to the fine-tuning of the initial conditions
at the level of ten percent:
$\left|\Delta t_{\rm ini}\right|/t_{\rm inf}\approx0.11$.
}
(its 3D shape close to the inflection point is very similar
to that shown in figure \ref{plot3d2}).
This kind of behavior is generic for models where the positivity of the
$\eta$ matrix trace is obtained due to the subleading corrections to the
K\"ahler potential.

\begin{table}[t]
\centering
		\begin{tabular}{|c|c|c|c|c|c|c|c|c|c|c|c|c|}
		\hline	$\Delta t_{\rm ini}$ & $-0.9$ & $-0.5$ &  $-0.3$ & $-0.2$ & $-0.1$ & $-0.01$ & $-0.001$ & $0$ & $0.001$ & $0.01$ & $0.03$  & $0.1$ \\ \hline
                $N$ & $220$   & $218$ &  $216$ &  $214$ & $205$   &  $139$ & $116$ &  $113$ & $110$  & $87$  & $52$ & $19$   \\ \hline
		\end{tabular}
\caption{Number of e-folds of inflation for several values of
$\Delta t_{\rm ini}= t_{\rm ini}-t_{\rm inf}$ in model
\ref{posthr}.}
	\label{tab3}
\end{table}

The fine-tuning of the parameters is similar to those in other models
with SUSY Minkowski minima (models \ref{posneg} and \ref{twopos}
discussed in the previous section).
The parameter $A$ has to be fine-tuned at the level
of $10^{-5}$ to obtain gravitino mass of the order of TeV. The flatness
of the potential, which is enough to provide $60$ e-folds of inflation,
is achieved due to the fine-tuning of the combination $cC_0/C_1$
at the level of $10^{-5}$.


\subsection{Negative exponent}
\label{negthr}

We now discuss a model with the superpotential (\ref{Wthr})
but for negative exponent: $c<0$.
In this case one cannot realize inflection point inflation that ends in the
SUSY Minkowski minimum. The reason is the same as for the KL model
of subsection \ref{KL}. The second minimum in the $t$-direction
(existing in addition to the SUSY Minkowski one) is supersymmetric,
has negative value of the potential and can not be deformed to an
inflection point with positive energy by any changes of the parameters
$A$, $C_0$ and $C_1$. Inflation can be realized if we add to
the potential (\ref{Vthr}) the uplifting term (\ref{uplift})
$\Delta V={E}/{t^2}$.
Similarly as in the KL model considered in subsection \ref{KL},
we use the K\"ahler potential in the leading approximation (\ref{leadK}).
The model is constructed in a similar fashion as the one proposed
in \cite{LiWe}. For some region of the parameter space the potential
(\ref{Vthr}) without uplifting (for $E=0$) has two  AdS local minima.
The minimum at larger volume is deeper than the one at smaller volume.
The uplifting term (\ref{uplift}) is used to lift both minima to dS space.
The value of $E$ is chosen in such a way that the minimum at larger volume
has positive but almost vanishing cosmological constant.
The minimum at smaller volume is more strongly lifted and acquires
a large cosmological constant. It is also more strongly lifted than
the local maximum separating both minima. So, for strong enough lifting,
this local maximum and the minimum at smaller volume may disappear.
With appropriate tuning of the parameters one can obtain a very flat
inflection (or saddle) point.
The position of such inflationary inflection point can be well
approximated by the formula for the position of the SUSY Minkowski
minimum $t_{\rm Mink}$ (\ref{minkthr}):
\begin{equation}
t_{\rm inf}\approx \frac{1}{|c|}-\frac{C_0}{C_1} \ .
\label{infthr}
\end{equation}
Parameters $C_0$ and $C_1$ should have different signs in order
to avoid too small, or even negative, $t_{\rm inf}$.

For the numerical example we choose the following set of parameters:
\begin{equation}
A=2.30689634\cdot10^{-7} , \hspace{0.5cm}
C_0=1\cdot10^{-4} , \hspace{0.5cm}
C_1=-2.5\cdot10^{-6} ,  \hspace{0.5cm}
c=-\frac{2\pi}{60} \,,\hspace{0.5cm}
E=1.7685\cdot10^{-16} .
\label{par4}
\end{equation}
The shape of the potential (\ref{Vthr}) before and after uplifting is
quite similar to that shown in figure \ref{plottKL}
for the KL model. The inflationary
inflection point, generated by the uplifting of the minimum at
smaller volume, is situated at $t_{\rm inf}\approx53.91$.
At that point the only non-vanishing entry of the $\eta$ matrix is
$\eta_{\tau}^{\tau}\approx115.7$.
The second slow roll parameter at the inflection point is very small:
$\epsilon\approx6\cdot10^{-15}$.
The (near) Minkowski vacuum is situated at $t_{\rm vac}\approx71.73$.
The corresponding gravitino mass is large, about $3\cdot10^8$ GeV.
The barrier, that separates the vacuum from
the run-away region, occurs at $t_{\rm bar}\approx101.76$. In this
model $\tau=0$ is a minimum in the $\tau$-direction for
any value of $t$ (the 3D shape of the potential in the region
important for inflation is very similar to that for the KL model
shown in figure \ref{plot3dKL}).

\begin{figure}[t!]
  \centering
  \includegraphics[width=10cm,angle=0]{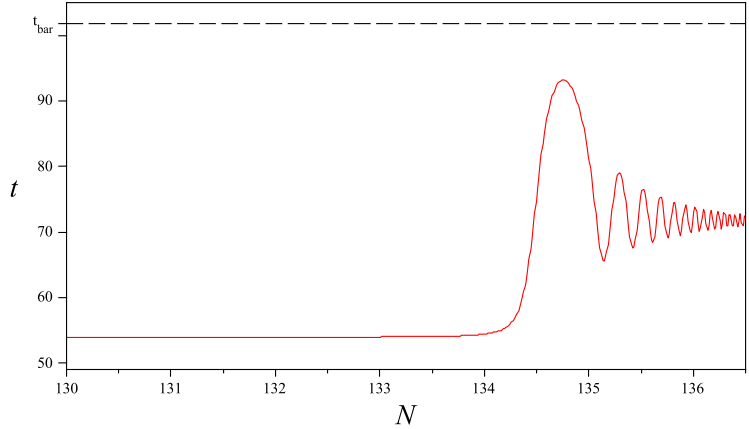}
  \caption{Evolution of the inflaton $t$ in the late stage of inflation
as a function of e-folds $N$ in model \ref{negthr}. The dashed line shows
the position of the barrier separating the vacuum at finite $t$ from the
run-away region.}
  \label{tevol4}
\end{figure}

Solving numerically the equations of motion (\ref{eqmt})-(\ref{eqmtau})
with the initial conditions $\tau(0)=\tau'(0)=t'(0)=0$, $t(0)=t_{\rm inf}$,
we obtain about 134 e-folds of inflation after which the inflaton
oscillates around the vacuum (see figure \ref{tevol4}).
Notice that due to Hubble damping the inflaton does not overshoot
the vacuum, even though the potential barrier is lower than
the height of the inflationary inflection point.
Table \ref{tab4} shows that, starting the evolution of the inflaton
from some region above the inflection point, one can obtain up to
about $260$ e-folds of inflation

\begin{table}
\centering
\begin{tabular}{|c|c|c|c|c|c|c|c|c|c|c|}
\hline	
$\Delta t_{\rm ini}$ & $-0.1$ & $-0.09$ &  $-0.08$ & $-0.06$ & $-0.01$ &
$-0.001$ & $0$ & $0.0001$ & $0.0005$ & $0.001$\\
\hline
$N$ & $14$  & $108$ &  $253$  &  $261$ & $262$ &
$233$ &  $134$ & $114$ & $61$  & $36$  \\
\hline
\end{tabular}
\caption{
Number of e-folds of inflation for several values of
$\Delta t_{\rm ini}= t_{\rm ini}-t_{\rm inf}$ in model
\ref{negthr}.}
	\label{tab4}
\end{table}

Let us now discuss fine-tuning of this model. In the presented example the
parameter $A$ has to be fine-tuned at the level of $10^{-8}$ to obtain at
least $60$ e-folds of inflation. However, it turns out that the degree of
fine-tuning depends on the ratio $|cC_0/C_1|$.
Stronger fine-tuning is needed for larger values of $|cC_0/C_1|$.
We recall that this ratio regulates also the
height of the barrier that separates the vacuum from the decompactification
region. Therefore, the fine-tuning required to obtain long lasting
inflation is strictly related to the height of this barrier. For example,
for $C_0=0$ the fine-tuning of $A$ at the level of $10^{-5}$ would be
enough to obtain more than $60$ e-folds of inflation.
However, as we mentioned before, for $C_0=0$ the barrier is too small
and the inflaton $t$ runs away to infinity. Numerical calculations show
that for the minimal height of the barrier necessary to protect the
inflaton from overshooting the vacuum, the fine-tuning of the parameter
$A$ is at the level of $10^{-8}$.

In the presented
example the fine-tuning of the initial conditions required to obtain more
than $60$ e-folds of inflation equals
$\left|\Delta t_{\rm ini}\right|/t_{\rm inf}\approx0.001$,
as seen from table \ref{tab4}. This fine-tuning could be relaxed
even by a few orders of magnitude for smaller
values of $|cC_0/C_1|$:
for example, $\left|\Delta t_{\rm ini}\right|/t_{\rm inf}\approx0.1$
for $C_0=0$. However, smaller values of  $|cC_0/C_1|$ are
not allowed because of the overshooting problem.
The above relations between the height of the barrier
and the strength of fine-tuning can be explained by the reasoning
presented at the end of subsection  \ref{KL}.


\section{Experimental constraints and signatures}
\label{exp}

We now compare predictions of our models to cosmological measurements.
First of all, we calculate the amplitude of density perturbations
\cite{Lyth}:
\begin{equation}
 \frac{\delta\rho}{\rho}=\frac{2}{5}\sqrt{{\cal P_R}(k_0)}
 \,,
\end{equation}
where $k_0\approx7.5H_0$ is the COBE normalization scale which
leaves the horizon approximately $55$ e-folds before the end of inflation
and ${\cal P_R}$ is the amplitude of the scalar perturbations
given, in the slow-roll approximation, by the following formula:
\begin{equation}
{\cal P_R}(k)=\frac{1}{24\pi^2}
\left.\left(\frac{V}{\epsilon}\right)\right|_{k=aH}
\,.
\end{equation}
In all models presented in this paper the parameters were chosen
in such a way as to generate the density perturbations with the
amplitude $\frac{\delta\rho}{\rho}\approx2\cdot10^{-5}$
consistent with COBE measurements.

\begin{figure}
  \centering
  \includegraphics[width=8cm,angle=0]{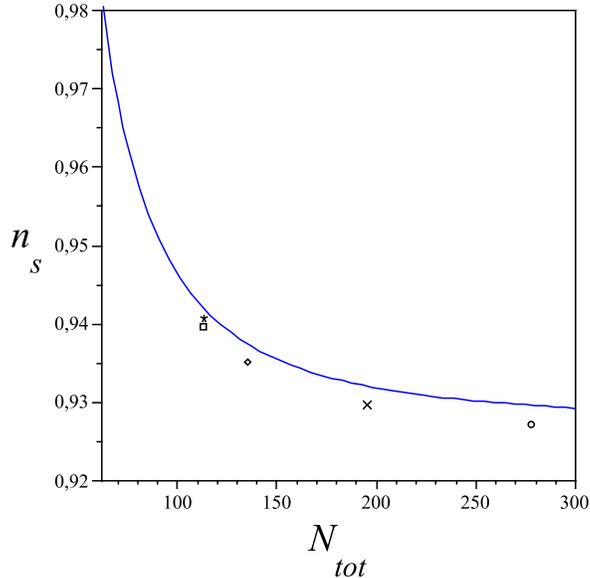}
  \caption{Comparison of the approximate analytic formula (\ref{fns}) for the
    spectral index at the COBE scale $N_e=55$ with the numerical results for
    different models with the initial conditions $t(0)=t_{\rm inf}$:  \small $\times$ \normalsize
    model \ref{posneg} with the parameters (\ref{par1}), $\circ$ model
    \ref{twopos} with the parameters (\ref{par2}), $\star$ model \ref{KL} with
    the parameters (\ref{parKL}),
\tiny $\square$ \normalsize
model \ref{posthr} with the parameters (\ref{par3}), $\diamond$
model \ref{negthr} with the parameters (\ref{par4}).}
  \label{nsLW}
\end{figure}

Secondly, we analyze the spectral index:
\begin{equation}
        n_s-1\equiv\frac{d\ln\mathcal{P}_{\mathcal{R}}(k)}{d\ln k}
\approx\frac{d\ln\mathcal{P}_{\mathcal{R}}(N)}{dN}\ ,
\end{equation}
where the last approximation comes from the fact that this quantity is
evaluated at horizon crossing $k=aH=He^N$ which implies $d\ln k\approx dN$.
Approximate analytical formula for the spectral index in models of
inflation starting from an inflection point was derived in
\cite{LiWe} (see also \cite{Baumann2,Sanchez,Itzhaki}):
\begin{equation}
\label{fns}
n_s\approx1
-\frac{2\pi}{N_{\rm tot}}\cot\left(\frac{\pi N_e}{2N_{\rm tot}}\right)
\,,
\end{equation}
where $N_{\rm tot}$ is the total number of e-folds during inflation
starting exactly at the inflection point while $N_e$ is the number
of e-folds between the time when a given scale crosses the horizon
and the end of inflation.
We are interested in the value of $n_s$ at the COBE normalization scale
corresponding to $N_e\approx55$ (the exact value of $N_e$ at the COBE
normalization scale depends on the reheating temperature which is model
dependent).
In figure \ref{nsLW} we show that the numerically found values of the
spectral index in all models presented in this paper are in a good
agreement with the approximate formula (\ref{fns}).
The 5-year WMAP result \cite{wmap}, $n_s=0.96\pm0.014$, slightly favours
models with smaller values of $N_{\rm tot}$ (with an exception of quite
unnatural situation when $N_{\rm tot}$ is very close to $N_{\rm e}$).
Models with very small slope of the potential at the inflection point
have large $N_{\rm tot}$ and the spectral index which can be more
than $2\sigma$ below the central WMAP value.
We checked numerically that the spectral index does not depend
significantly on the initial conditions. This is rather intuitive
because, when the inflaton starts at some very flat region above
the inflection point, it approaches that inflection point with a very
small velocity. Therefore, the time dependence of the inflaton field
below the inflection point is almost the same as in the case when
it starts at rest exactly at the inflection point, and the COBE
normalization scale corresponds to almost the same value of the
inflaton.

For each model considered in this paper there is a region of the
corresponding parameter space for which the period of inflation
starting exactly at the inflection point is shorter than $55$ e-folds
while it is longer than $55$ e-folds when the inflaton starts
above the inflection point.  However, such parameters give the spectral
index at the COBE scale bigger than unity. This follows from the
fact that in such cases the COBE scale crosses the horizon when
the inflaton is above the inflection point where the parameter
$\eta$ (which determines the value of the spectral index) is positive.
Therefore, in order to be consistent with the data, at least $55$ e-folds
have to be generated when the inflaton is below the inflection point.

The tensor to scalar ratio depends mainly on the value of the
slow-roll parameter $\epsilon$ at the inflationary inflection point,
$r\approx16\epsilon$. So, in models with the SUSY (near)
Minkowski minimum $r$ is bigger than in more fine-tuned models with
uplifting. Nevertheless, in all models discussed in this paper,
as is typical for string-inspired inflation\footnote
{
See, however, \cite{Silv} and \cite{Cicoli} for recent proposals for
generating observable gravity waves within string theory.
},
the value of $r$ is many orders of magnitude
below the sensitivity of near future experiments.
We found also that the running of the spectral index
$\frac{d n_s}{d\ln k}$ in inflection point inflation is typically of
the order of $10^{-3}$ so it is also negligible from the observational
point of view.


\section{Fine tuning and time dependent potentials}
\label{timedepend}

The authors of \cite{Itzhaki} postulate the existence of a
time-dependent part of the potential, that is an increasing function
of the volume modulus $t$, as a possible solution to the problem of
initial conditions for the inflaton in models of inflection point
inflation. They argue that this kind of potential can
be generated in higher dimensional models by (0+p)-branes which
are point-like objects from the 4D point of view and wrap p-cycles
in the internal
manifold\footnote
{
This scenario is even more attractive because, as it was argued
in \cite{Itzhaki2}, it leads to formation of giant spherically
symmetric overdense regions in the observed part of the Universe
which in principle could be detected.
}.
Using the number of e-folds $N$ instead of the cosmic time we can write
such correction to the potential in the following form:
\begin{equation}
\label{pbrane}
\Delta V_{0+p}=kn_{0}t^{p-3}e^{-3N} \,,
\end{equation}
where $k$ is a constant of order one and $n_0$ is the initial density of
(0+p)-branes. For $p>3$ this is indeed an increasing function of the
volume modulus $t$. The time-dependence of this potential is due to the
expansion of the Universe which decreases the matter density.
If the initial density of (0+p)-branes is large enough, the minimum
of the full potential is located at some value of $t$ smaller than
$t_{\rm inf}$.
We want to solve the initial conditions problem, so we do not assume
that the initial value of the inflaton is very close to that minimum.
Thus, for early times the inflaton oscillates around a time-dependent
minimum of the full potential. Two things happen for increasing time.
First: the minimum of the full potential moves towards larger values
of $t$. Second: the amplitude of the inflaton oscillations decreases
due to the Hubble friction.
If the initial value of the inflaton field is not too far away from
the initial position of the minimum of the full potential, those
oscillations may be damped very strongly before the minimum of the
full potential moves to $t$ larger than the position of the inflection
point of the time-independent part of the potential, $t_{\rm inf}$.
In such a case the volume modulus moves together with the minimum
of the full potential towards larger values until it reaches a position
very close to the inflection point where it stops tracking the
time-dependent minimum. Then the usual period of slow-roll inflation begins.

This mechanism does not solve the problem of the initial conditions
in models with SUSY (near) Minkowski minima because typically the
potential becomes unstable in the $\tau$-direction if we move too
far above the inflection point.
On the other hand, it works very well in models with uplifting.
It is shown in tables \ref{tabn1} and \ref{tabn2} that in both models
with uplifting presented in this paper even small amount of the initial
density of (0+p)-branes is sufficient to solve the initial
conditions problem.

In \cite{Itzhaki} it was argued that the potential (\ref{pbrane}) can
also significantly reduce the fine-tuning of the parameters that are
responsible for the flatness of the potential. The argument
used
by the authors of \cite{Itzhaki} was that one can obtain more than
$60$ e-folds of inflation even if the inflection point is not flat
at all. However, in what follows we show that such case is ruled out
by the present experimental constraints on the spectral index.

\begin{table}
\begin{minipage}[t]{0.48\linewidth}
\centering
\begin{tabular}{|c|c|c|c|c|c|}
\hline	
$n_0$ & $10^{-25}$ & $10^{-24}$ &  $10^{-20}$ & $10^{-18}$ & $10^{-17}$
\\
\hline
$t_1$ & $53$  & $49.5$ &  $14$  &  $2.4$ & $0.9$
\\ \hline
$t_2$ & $55.4$  & $57.1$ &  $788$  &  $6700$ & $35000$
\\ \hline
\end{tabular}
\caption{
The range $t_1<t_{\rm ini}<t_2$ of the initial conditions for the
inflaton which provide more than $60$ e-folds of inflation for different
amount of initial density of (0+p)-branes in model \ref{KL} with the
parameters (\ref{parKL}). Time dependent potential (\ref{pbrane}) is
used with $k=1$ and $p=6$.}
\label{tabn1}
\end{minipage}
\hfill
\begin{minipage}[t]{0.48\linewidth}
\centering
\begin{tabular}{|c|c|c|c|c|c|}
\hline	
$n_0$ & $10^{-25}$ & $10^{-24}$ &  $10^{-20}$ & $10^{-18}$ & $10^{-17}$
\\ \hline
$t_1$ & $52.4$  & $49.4$ &  $13$  &  $2$ & $0.8$
\\ \hline
$t_2$ & $53.8$  & $53.7$ &  $510$  &  $3900$ & $21000$
\\ \hline
		\end{tabular}
	\caption{
The same as in table \ref{tabn1} but for model \ref{negthr} with the
parameters (\ref{par4})}
	\label{tabn2}
\end{minipage}
\end{table}

When the inflection point is not fine-tuned to be flat, inflation may
take place only when the inflaton moves together with the minimum of
the full potential. In order to obtain more than $60$ e-folds of this
inflationary phase the inflection point and the Minkowski vacuum have
to be situated at very large values of $t$. Moreover, the Hubble
parameter decreases significantly during the evolution of the inflaton,
so this is not a standard quasi-exponential inflation but rather a power
law one:
\begin{equation}
a\propto t_{\rm cosm}^{\beta} \,.
\end{equation}
It can be easily shown that the power exponent $\beta$ is related
to the Hubble parameter:
\begin{equation}
\label{betaH}
\beta=-\frac{H}{H'} \,,
\end{equation}
where prime denotes derivative with respect to number of e-folds $N$.
In power law inflation the spectral index is given by
\cite{LySt}:
\begin{equation}
\label{nsbeta}
n_s=1-\frac{2}{\beta-1}
\,.
\end{equation}
In order to be compatible with the data $\beta$ should be rather large.
For example $n_s>0.9$ requires $\beta>21$.
In order to estimate $\beta$, we consider a toy-model with the following
potential:
\begin{equation}
\label{pottoy}
V=\frac{F}{t^q}+Gt^se^{-3N} \,.
\end{equation}
The first term simulates a time-independent part of the potential.
The second term is the time-dependent potential (\ref{pbrane}) with
$G\equiv kn_0$ and $s\equiv p-3$. The position of a time-dependent
minimum $t_{\star}$ reads:
\begin{equation}
t_{\star}=\left(\frac{qF}{sG}\right)^{1/(q+s)}e^{3N/(q+s)} \,.
\end{equation}
Therefore, the value of the potential (\ref{pottoy}) at a
time-dependent minimum is given by:
\begin{equation}
\label{potstar}
V(t_{\star})=
\left[\left(\frac{q}{s}\right)^{s/(q+s)}
+\left(\frac{s}{q}\right)^{q/(q+s)}\right]
F^{s/(q+s)}G^{q/(q+s)}e^{-(3qN)/(q+s)}
\,.
\end{equation}
In order to find the power exponent $\beta$ one needs $H'/H$
(see eq.\ (\ref{betaH})). Assuming that the Hubble parameter changes
mainly due to the change of the potential energy of the inflaton and
not due to the change of its kinetic energy, the following relation
holds:
\begin{equation}
\frac{H'}{H}\approx\frac{V'}{2V} \,.
\end{equation}
Hence, using (\ref{betaH}) and (\ref{potstar}), we obtain the
following exponent of power law inflation
\begin{equation}
\label{betaqs}
\beta\approx\frac{2}{3}\left(1+\frac{s}{q}\right)\,.
\end{equation}
We have calculated numerically exponent $\beta$ for several
values of parameters $q$ and $s$.
The results are in a very good agreement with the above approximate
formula.
The tree-level potentials as well as the uplifting terms in
models \ref{KL} and \ref{negthr} behave like $t^{-2}$ for small $t$.
Therefore, in those models exponent $\beta$
equals approximately $1$, $4/3$ or $5/3$ for $p=4$, $p=5$ or $p=6$,
respectively. We confirmed these results by our numerical analysis.
Such values of the exponent are too small to give
a realistic model of power law inflation (for $p=4$ there is no
inflation at all because in this case $\ddot a=0$.)
There are two reasons: First,
inflation with so small exponents should be very long in order to
produce at least 60 e-folds. This could be achieved only for a
very unnatural choice of the parameters. Second, such small
values of $\beta$ lead to values of the spectral index well outside
experimental bounds.

Using eqs.\ (\ref{nsbeta}) and (\ref{betaqs}) we find that
the spectral index for inflation driven by potential
(\ref{pottoy}) reads
\begin{equation}
n_s\approx1-\frac{6}{2s/q-1}
\,.
\label{nsqs}
\end{equation}
It can be in the range $0.9<n_s<1$ only if $s\gtrsim30q$.
In the mechanism proposed in \cite{Itzhaki} the biggest
possible value of $s$ equals 3 (for $p=6$).
The period of inflation with the inflaton tracking
the time-dependent minimum of the potential gives much
too small values of the spectral index. Thus, the COBE normalization
scale must correspond to that part of inflation when
the inflaton is close to a very flat inflection point.
We conclude that the mechanism proposed in \cite{Itzhaki}
can not solve the problem of fine-tuning of the parameters
of the potential. It can solve only the problem of the fine-tuning
of the initial condition.


\section{Discussion and conclusions}
\label{discussion}

We considered five models in which the volume modulus
plays the role of the inflaton. In each of them inflation takes place
in the vicinity of an appropriately flat inflection point of the
potential. Those models may be divided into two classes. In the first
class (models \ref{posneg}, \ref{twopos} and \ref{posthr}) inflation
ends in a SUSY (near) Minkowski minimum.
In this class of models the corrections to the K\"ahler potential
are necessary to realize inflation. If a SUSY (near) Minkowski
minimum occurs at smaller volume than an inflationary inflection
point (as in model \ref{posneg}), larger amount of corrections
are required than in the opposite case (models \ref{twopos}
and \ref{posthr}). Nevertheless, in all models the corrections
are small enough to trust the perturbative expansion of the K\"ahler
potential. In the second class (models \ref{KL} and \ref{negthr})
inflation ends in a vacuum obtained by uplifting
of some deep AdS SUSY minimum.
The main phenomenological difference between those classes is
the relation between the gravitino mass and the Hubble constant during
inflation.
In the first class the gravitino mass may be orders of magnitude
smaller while in the second class it is bigger than the Hubble constant.
So, only in models of the first class a light gravitino (e.g.\ with mass
in the TeV range) is compatible with high scale (e.g.\ GUT scale) inflation.

The structure of the superpotential decides to which class a given model
belongs. The first class contains models in which at least one
gaugino condensation term has a positive exponent. Models with only
negative exponents, like e.g.\ the KL-type models, are in the second class
and can not accommodate the gravitino mass much smaller than the
inflationary Hubble constant.

From a technical point of view, the crucial difference between the models
from different classes is the  structure of the potential. In each case
the parameters may be chosen in such a way that the potential has a
SUSY Minkowski minimum. The potential has (at least for some range
of parameters) also the second minimum. This second minimum is
supersymmetric in models with only negative exponents in the nonperturbative
terms in the superpotential.
It is non-supersymmetric when at least one of the gaugino condensation
term has a positive exponent. Only those non-supersymmetric minima
may be deformed into inflection points with a positive value of the
potential, keeping the SUSY Minkowski minimum intact.

There is another interesting difference between the two classes of models.
All models require some amount of fine-tuning of the parameters and
the initial conditions to produce long enough 
inflation\footnote
{The fine-tuning of the parameters in the context of inflection point 
brane-antibrane inflation  \cite{Baumann} was analyzed in \cite{Hoi}. 
The problem of initial conditions in that context was investigated 
in \cite{Underwood}
}. 
However, models with only negative exponents (and uplifting) need 
much stronger fine-tuning than models with at least one positive 
exponent (and no uplifting). In the latter models,
the fine-tuning of the parameters is typically at the level
of $10^{-5}$. The fine-tuning of the initial conditions for the
inflaton field is at the level of a few percent. Models with only
negative exponents require much stronger fine-tuning: about $10^{-8}$
for the parameters and $10^{-3}$ for the initial conditions.
Such substantial differences in fine-tuning are related to the
overshooting problem. In models with only negative exponents,
the barrier which protects from overshooting the vacuum is generated
by the uplifting. Typically the uplifting in the region of the barrier is
smaller than in the region of the vacuum or the inflection point.
As a result the height of the barrier is much smaller than the height
of the inflection point at which inflation starts. Additional
fine-tuning of the parameters (additional factor of order $10^{-3}$)
is necessary to get a high enough barrier. On the other hand,
in models of the first class, a term in the superpotential with
a positive exponent automatically produces a high barrier.

The difference in the fine-tuning of the initial conditions is not
as large as the difference in the fine-tuning of the superpotential
parameters. The initial conditions in models with only negative
exponents must be fine-tuned (only) about 10 times stronger than
in models with at least one positive exponent. Moreover, only
in models with all negative exponents the initial conditions problem
can be solved by the mechanism proposed in \cite{Itzhaki}.

Finally, we should also mention that technically it is possible
to construct inflationary models ending in the SUSY (near) Minkowski
minimum with single gaugino condensate and a negative exponent
if we allow for much more complicated function of $T$ which stands
in front of the exponential term in the superpotential (e.g.\
containing 4-th or higher powers of $T$). However,
we do not discuss this kind of models in detail since they seem
to be quite unnatural.

To summarize let us recall briefly advantages and disadvantages of
both classes of models of inflection point inflation.
Models with only negative exponents in the nonperturbative terms
in the superpotential require strong uplifting and so have the
gravitino mass larger than the Hubble constant during inflation.
The fine-tuning of parameters in those models is about 3 orders
of magnitude stronger than in other models. The fine-tuning
of the initial conditions is also stronger but this problem
can be solved by appropriate time dependent contributions to the
potential. Models with at least one positive exponent in
gaugino condensation terms do not require strong uplifting
and as a result may accommodate the gravitino several orders of magnitude
lighter than the inflationary Hubble scale. The fine-tuning of the
parameters is typical for inflation (and much weaker than
in models with only negative exponents). The necessary fine-tuning
of the initial conditions is at the level of a few percent but
can not be made weaker by the mechanism of \cite{Itzhaki}.
Corrections to the lowest order K\"ahler potential are necessary
in those models in order to have an inflection point stable
in the axion direction. The size of such corrections must be
bigger than some minimal value but one can argue that they are still
reasonably small.


\section*{Acknowledgments}

Work partially supported by the UE 6th Framework Programs
MRTN-CT-2004-503369 ``Quest for Unification''
and MTKD-CT-2005-029466
``Particle Physics and Cosmology: the Interface''.

\end{document}